\documentclass[aip, amsmath, amssymb, reprint, floatfix]{revtex4-1}

\usepackage{dcolumn}
\usepackage{bm}
\usepackage[english]{babel}
\usepackage[utf8]{inputenc}
\usepackage[T1]{fontenc}
\usepackage{mathptmx}
\usepackage{etoolbox}
\usepackage[caption=false]{subfig}
\usepackage{graphicx}
\usepackage[colorlinks=true,
            linkcolor=blue,
            urlcolor=blue,
            citecolor=blue]{hyperref}
\usepackage[usenames,dvipsnames]{color}
\usepackage[version=3]{mhchem}
\usepackage[textsize=small]{todonotes}
\usepackage{graphicx}
\usepackage{textcomp}
\usepackage{float}

\allowdisplaybreaks

\newcommand{\cref}{c^{\ominus}}
\newcommand{\muref}{\mu^{\ominus}}

\makeatletter
\def\@email#1#2{%
 \endgroup
 \patchcmd{\titleblock@produce}
  {\frontmatter@RRAPformat}
  {\frontmatter@RRAPformat{\produce@RRAP{*#1\href{mailto:#2}{#2}}}\frontmatter@RRAPformat}
  {}{}
}%
\makeatother

\begin{document}

\preprint{APS/123-QED}

\title{A Generalized Grand-Reaction Method for Modelling the Exchange of Weak (Polyprotic) Acids between a Solution and a Weak Polyelectrolyte Phase}

\author{David Beyer}
\affiliation{Institute for Computational Physics, University of Stuttgart, D-70569 Stuttgart, Germany}
\author{Christian Holm}%
\affiliation{Institute for Computational Physics, University of Stuttgart, D-70569 Stuttgart, Germany}
\email{holm@icp.uni-stuttgart.de}

\date{\today}

\begin{abstract}
We introduce a Monte-Carlo method that allows for the simulation of a polymeric phase containing a weak polyelectrolyte, which is coupled to a reservoir at a fixed pH, salt concentration and total concentration of a weak polyprotic acid. 
The method generalizes the established Grand-Reaction Method by Landsgesell et al. [Macromolecules \textbf{53}, 3007-3020 (2020)] and thus allows for the simulation of polyelectrolyte systems coupled to reservoirs with a more complex chemical composition.
In order to set the required input parameters that correspond to a desired reservoir composition, we propose a generalization of the recently published chemical potential tuning algorithm of Miles et al. [Phys. Rev. E \textbf{105}, 045311 (2022)].
To test the proposed tuning procedure, we perform extensive numerical tests for both ideal and interacting systems.
Finally, as a showcase, we apply the method to a simple test system which consists of a weak polybase solution that is coupled to a reservoir containing a small diprotic acid.
The complex interplay of the ionization various species, the electrostatic interactions and the partitioning of small ions leads to a non-monotonous, stepwise swelling behaviour of the weak polybase chains.
\end{abstract}

\maketitle

\section{Introduction}
Electrically charged polymers, commonly called ``polyelectrolytes'', are a versatile class of materials with many applications and interesting properties.
Simple polyelectrolyte chains can for instance be used as thickening agents in hygiene products such as shampoo or as flocculants in water treatment.
More complex polyelectrolyte architectures, such as polyelectrolyte networks (``hydrogels'') allow for even more sophisticated applications.
Hydrogels are for instance used in areas as diverse as medicine,\cite{drury03a} agriculture,\cite{kazanskii92a} for desalination\cite{hoepfner10a, hoepfner13b, richter17a, arens19a, rud21a} and hygiene products.\cite{masuda94a}
Other polyelectrolyte architectures include polyelectrolyte brushes,\cite{chen17a, xu19a, ferranddrake20a} which can for instance be used for protein purification and for the stabilization of colloidal solutions, and polyelectrolyte coarcervates.\cite{sing2020a, sing2017a, li2018a, perry2014a, chang2017a, fu2017a}
In addition to many applications, polyelectrolytes are also of fundamental interest to molecular biology and biochemistry, since many biological macromolecules, such as proteins, DNA and RNA are in fact polyelectrolytes.\cite{berg02a}
The presence of long-range electrostatic interactions makes the modelling of polyelectrolytes challenging from the point of view of theoretical and computational soft matter physics.
For instance, special techniques are needed to deal with these interactions in an efficient way in computer simulations.
In many cases, for example in many proteins,\cite{lund05a} the modelling can be even further complicated by the presence of some kind of association-dissociation reaction which leads to a complicated coupling between the chemical equilibrium, electrostatic interactions and the conformational degrees of freedom.\cite{landsgesell19a}
A paradigmatic example for such ``weak polyelectrolytes'' is a weak polyacid, i.e. a polymer chain consisting of weak acid monomers \ce{HA}, which can become charged by releasing a proton into solution:
\begin{align}
 \ce{HA <=> A- + H+}.
\end{align}
Ideal weak acid particles, for instance realized experimentally in dilute solutions of individual weak acid molecules (i.e. no chains), are well-described by the Henderson-Hasselbalch equation,
\begin{align} 
\alpha = \frac{1}{1+10^{\text{p}K_\text{A}-\text{pH}}}, 
\label{eq:Henderson_Hasselbalch}
\end{align}
which relates the average degree of ionization $\alpha$ with the pH of the solution and the p$K_\text{A}$-value of the considered molecules.
In contrast to this simple case, the ionization behaviour of weak polyacids is strongly altered by the electrostatic interactions of ionized monomers, especially by the electrostatic repulsion of neighbouring monomers.
These non-idealities lead to strongly shifted and deformed ionization curves as compared to the ideal theory.
Because the ionization behaviour and the conformational degrees of freedom of weak polyelectrolytes are coupled in a complicated way, this effect, which has been termed the ``polyelectrolyte effect'' in the past,\cite{landsgesell20b} is difficult to describe analytically.
Consequently, a theoretical interest in weak polyelectrolytes has led to the development of several numerical methods\cite{reed92a, smith94c, johnson94a, labbez07b, landsgesell17a, landsgesell20b, curk22a} over the decades in order to treat this problem.

Some time ago, Landsgesell et al.\cite{landsgesell20b} introduced the Grand-Reaction Monte-Carlo (G-RxMC) method.
This algorithm allows for the simulation of two-phase systems consisting of a solution containing small ions and a polymeric phase containing a weak polyacid and/or polybase (in addition to the small ions).
The method uses Monte-Carlo moves to model the exchange of small ions between the phases as well as the acid-base reactions in the polymeric phase. 
In contrast to earlier methods like the Reaction-Ensemble\cite{smith94c, johnson94a} or constant-pH method,\cite{reed92a} this new method is applicable over the whole range of pH-values.
Furthermore, in addition to the aforementioned ``polyelectrolyte effect'' it also correctly models the Donnan partitioning of ions between the polymeric phase and the solution, which in addition to the "polyelectrolyte effect" also influences the ionization behaviour.
By accounting for both the charge regulation as well as the Donnan partitioning, this new method has, for the first time, made particle-based simulations of weak polyelectrolyte hydrogels, which are a natural realization of such a two-phase system, possible.\cite{landsgesell22a}

One of the current limitations of the G-RxMC method in its original form is that the composition of the considered reservoir is fairly simple, as it contains only monovalent ions. 
For instance, it does not allow for the exchange of weak ions (e.g. small weak acid molecules or even pH-responsive chain molecules such as polypeptides) between the two phases.
In this publication, we show how the G-RxMC method can be extended to also model the exchange of small weak acid molecules.
In particular, we also demonstrate how the chemical potentials can be dynamically tuned to achieve the desired reservoir composition.
Since the reservoir composition is given in terms of concentrations, but the method takes as its input parameters chemical potentials, this problem is in fact non-trivial and could not be adequately addressed before.

\section{Setup and Method}

\begin{figure*}[htb]
\centering
\includegraphics[width=0.9\textwidth]{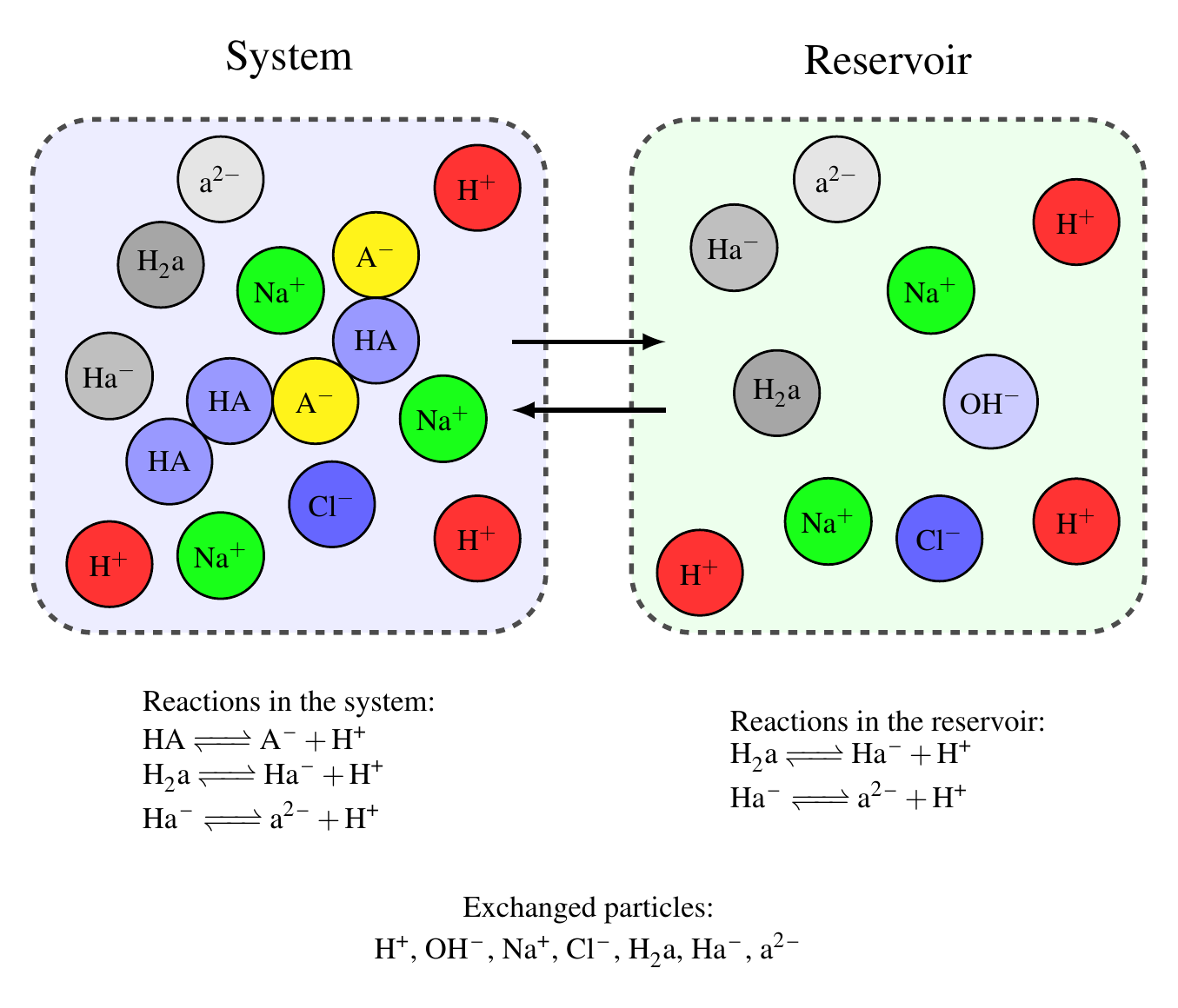}
\caption{\label{fig:schematic} Schematic representation of the setup considered in this publication. The two-phase setup consists of a polymeric phase (``system'') and an aqueous solution (``reservoir''). The reservoir has a fixed pH value and contains salt and small $n$-protic acid molecules (shown here: diprotic $\ce{H2a}$) at fixed concentrations. In addition to these constituents, the system also contains a weak polyelectrolyte, here shown as a weak polyacid with monomers $\ce{HA}$. All species except the weak polyacid can be exchanged between the two phases.
The exchange moves have to conserve the electroneutrality.}
\end{figure*}

In this paper, we will always assume a coarse-grained representation which explicitly models the polymer chains and ions, but treats the solvent only implicitly.
We note that an extension of the presented method to atomistic simulation models would be non-trivial, since the p$K_{\mathrm{a}}$-values of the various groups cannot be an input to an atomistic simulation but should in fact be an output.
Furthermore, it is important to note that in our simulations, each ion type has a distinct label, even if they are physically (i.e. with regards to their interactions) indistinguishable. 
As shown recently by Curk et al.\cite{curk22a} in the context of the standard G-RxMC method, one could in principle also employ unified ion types, reducing the number of distinct chemical reactions. 
However, because the concept of unified ion types is incompatible with the $\mu$-tuning scheme proposed in the next section, we refrain from using it in this publication. 
Still, using the equilibrium constants obtained from the $\mu$-tuning method, one could in principle employ unified ions in the actual simulation of the system.

To specify the setup, as shown schematically in \autoref{fig:schematic}, we consider the total system to consist of two distinct phases:
\begin{enumerate}
 \item A polymeric phase (e.g. a hydrogel or a coacervate), called in the following the ``system'', containing a weak polyacid characterized by p$K_\mathrm{A}$ with monomers \ce{HA} (neutral) and \ce{A-} (ionized) as well as small ions (\ce{H+}, \ce{OH-}, \ce{Na+}, \ce{Cl-}). 
 Furthermore, there is a small weak $n$-protic acid $\ce{H_$n$ a}$ (neutral), characterized by $n$ p$K_\mathrm{a}$-values p$K_\mathrm{a}^1$, p$K_\mathrm{a}^2$, ..., p$K_\mathrm{a}^n$, which can become ionized (\ce{H$_{n-1}$a-}, \ce{H$_{n-2}$a^{2-}}, ..., \ce{a$^{n-}$}).
 Both $\ce{H_$n$ a}$ and all of its ionized states can also be exchanged with the reservoir.
 \item An aqueous solution (``reservoir'') containing small ions (\ce{H+}, \ce{OH-}, \ce{Na+}, \ce{Cl-}) at fixed values of pH and $c_{\mathrm{NaCl}}^\mathrm{res}$ and the small weak $n$-protic acid $\ce{H_$n$ a}$ at a fixed total concentration $c_{\ce{H_$n$ a}}^{\mathrm{res},0}=\sum_{z=0}^{n}c_{\ce{H$_{n-z}$a$^{z-}$}}^{\mathrm{res}}$, i.e. the total dissolved amount of acid divided by the system volume. Note that while the total amount of acid is a free parameter that is needed to characterize the composition of the reservoir, the ratio of the different ionization states is fully determined by the chemical equilibrium.
\end{enumerate}
The phases are coupled grand-canonically, i.e. they have the same inverse temperature $\beta$ and electrochemical potentials $\hat{\mu}_i=\mu_i+z_i\psi^\text{Don}$,
where $\psi^\text{Don}$ is the Donnan potential and $i=\ce{H+},$ \ce{OH-}, \ce{Na+}, \ce{Cl-}, \ce{H_$n$ a}, \ce{H$_{n-1}$a-}, \ce{H$_{n-2}$a^{2-}}, ..., \ce{a$^{n-}$}.
The phases are in an electrochemical rather than a simple chemical equilibrium due to the macroscopic electroneutrality constraint imposed on both phases.\cite{landsgesell20b}
Although the considered systems are finite, the electroneutrality constraint still holds since both simulation boxes are supposed to represent typical subsystems of the macroscopic phases.
This implies that the system and the reservoir can only exchange pairs of small ions (when only monovalent ions are involved) or more generally groups of $z+1$ small ions (when a multivalent ion of valency $z$ is involved), rather than individual ion particles, which would violate the electroneutrality.
In this approach, the Donnan potential is a quantity that emerges automatically and does not need to be put in "by hand". 
As shown in our earlier publications,\cite{landsgesell20b, landsgesell22a, beyer22a} the Donnan potential can be determined a-posteriori from the simulation.
Formally, we represent the insertion and deletion moves by a set of virtual chemical reactions.
The following four reactions with the indicated equilibrium constants $K_i$ are always present (compare Landsgesell et al.\cite{landsgesell20b}):
\begin{align}
\emptyset \ce{<=> &H+ + OH-},\qquad &K_{\ce{H+},\ce{OH-}}\label{eq:insertion_water}\\
\emptyset \ce{<=> &Na+ + Cl-},\qquad &K_{\ce{Na+},\ce{Cl-}}\label{eq:insertion_nacl}\\
\emptyset \ce{<=> &Na+ + OH-},\qquad &K_{\ce{Na+},\ce{OH-}}\label{eq:insertion_naoh}\\
\emptyset \ce{<=> &H+ + Cl-},\qquad &K_{\ce{H+},\ce{Cl-}}\label{eq:insertion_hcl}.
\end{align}
Although the reactions are described by a total of four equilibrium constants, only two of them are independent parameters since $K_{\ce{H+},\ce{OH-}}=10^{-14}$ is fixed as the ionic product of water and 
\begin{align}
K_{\ce{Na+},\ce{OH-}} = \frac{K_{\ce{Na+},\ce{Cl-}}K_{\ce{H+},\ce{OH-}}}{K_{\ce{H+},\ce{Cl-}}}.
\end{align}
The insertion and deletion moves involving \ce{H_$n$ a} and its ionized forms can in the most general form be written as
\begin{align}
\emptyset \ce{<=> &$(z-l)$ H+ + $l$ Na+ + \ce{H$_{n-z}$a$^{z-}$}}
\end{align}
with the equilibrium constant
\begin{align}
K_{(z-l)\ce{H+},\,l\ce{Na+},\,\ce{H$_{n-z}$a$^{z-}$}}
\end{align}
where $l=0,...,z$ counts the number of \ce{Na+} ions involved in the reaction. 
Consequently, there are $z+1$ insertion and deletion moves involving \ce{H$_{n-z}$a$^{z-}$} and a total of $4+\sum_{z=0}^{n}(z+1) = 4 + (n+1)(n+2)/2$ insertion and deletion moves (including \autoref{eq:insertion_water}, \autoref{eq:insertion_nacl}, \autoref{eq:insertion_naoh} and \autoref{eq:insertion_hcl}).
The equilibrium constant $K_{\ce{H$_{n}$a}}$ for the insertion of the neutral species is simply determined by the chemical potential $\mu_{\ce{H_$n$ a}}$.
For the insertion reactions involving charged species, the reaction constants can in general be written as
\begin{align}
K_{(z-l)\ce{H+},\,l\ce{Na+},\,\ce{H$_{n-z}$a$^{z-}$}} = \left(\prod_{i=1}^{z}K_{\ce{a}}^i\right)\left(\frac{K_{\ce{Na+},\ce{Cl-}}}{K_{\ce{H+},\ce{Cl-}}}\right)^l K_{\ce{H$_{n}$a}},
\end{align}
where $K_{\ce{a}}^i$ is the equilibrium constant associated with the dissociation reaction of $\ce{H$_{n-(i-1)}$a$^{(i-1)-}$}$.
This means that the equilibrium constants $K_{(z-l)\ce{H+},\,l\ce{Na+},\,\ce{H$_{n-z}$a$^{z-}$}}$ are completely determined by the other equilibrium constants.
In addition to the insertion reactions, each of the $n$ not fully ionized species \ce{H$_{n-z}$a$^{z-}$} ($z=0,...,n-1$) can dissociate in a reaction of the form
\begin{align}
\ce{H$_{n-z}$a$^{z-}$ <=> & H+ + H$_{n-(z+1)}$a$^{(z+1)-}$}
\end{align}
where the equilibrium constant 
\begin{align}
 K_{\ce{a}}^{z+1} = 10^{-\mathrm{p}K_{\ce{a}}^{z+1}}
\end{align}
is determined by the specific chemistry of the small acid particles and thus and input parameter in our coarse-grained simulation setup.
To avoid sampling bottlenecks, we also include the following reformulations of the dissociation reaction:\cite{landsgesell20b}
\begin{align}
\ce{H$_{n-z}$a$^{z-}$ <=> & Na+ + H$_{n-(z+1)}$a$^{(z+1)-}$}\\
\ce{H$_{n-z}$a$^{z-}$ + OH- <=> & H$_{n-(z+1)}$a$^{(z+1)-}$}\\
\ce{H$_{n-z}$a$^{z-}$ + Cl- <=> & H$_{n-(z+1)}$a$^{(z+1)-}$}
\end{align}
with equilibrium constants
\begin{align}
K_{\ce{a}}'^{z+1} &= K_{\ce{a}}^{z+1}\frac{K_{\ce{Na+},\ce{Cl-}}}{K_{\ce{H+},\ce{Cl-}}}\\
K_{\ce{a}}''^{z+1} &= \frac{K_{\ce{a}}^{z+1}}{K_{\ce{H+},\ce{OH-}}}\\
K_{\ce{a}}'''^{z+1} &= \frac{K_{\ce{a}}^{z+1}}{K_{\ce{H+},\ce{Cl-}}}.
\end{align}
Finally, inside the polymeric phase there is also the dissociation reaction of the weak polyacid HA:
\begin{align}
\ce{HA <=> A- + H+},\,\qquad K_\mathrm{A}
\label{eq:acid_ionization_grxmc}
\end{align}
which is described by the equilibrium constant $K_\mathrm{A}$. 
As before, one should also include the following linear combinations in order to avoid sampling bottlenecks:
\begin{align}
\ce{HA <=> &A- + Na+}\\
\ce{HA + Cl- <=> &A-}\\
\ce{HA + OH- <=> &A-}
\end{align}
with equilibrium constants
\begin{align}
K_{\ce{A}}' &= K_{\ce{A}}\frac{K_{\ce{Na+},\ce{Cl-}}}{K_{\ce{H+},\ce{Cl-}}}\\
K_{\ce{A}}'' &= \frac{K_{\ce{A}}}{K_{\ce{H+},\ce{OH-}}}\\
K_{\ce{A}}''' &= \frac{K_{\ce{A}}}{K_{\ce{H+},\ce{Cl-}}}.
\end{align}
Furthermore, one needs to include the dissociation reactions
\begin{align}
\ce{H$_{n-z}$a$^{z-}$ + $z\cdot$HA <=> $z\cdot$A-}
\end{align}
with the equilibrium constants
\begin{align}
	\tilde{K}_z = \frac{\left(K_{\ce{A}}\right)^z}{K_{\ce{H$_{n}$a}}\prod_{i=1}^{z}K_{\ce{a}}^i}
\end{align}
These linear combinations become important when the system contains mostly multivalent ions.

From a theoretical point of view, the presented set of reactions is redundant, however it guarantees a thorough sampling in a simulation.
It would be straightforward to generalize the described approach to a reservoir containing different polyprotic acids and a system containing one (or multiple) polyprotic polyacids. 
Also, the whole framework is not only applicable to weak acids but also to weak bases.
Since these generalizations only clutter the notation with even more indices but do not add any fundamentally new challenge, we will in the following focus on the case outlined above.

Given the set of chemical reactions, it is straightforward to implement them using the well-established Reaction-Ensemble Monte-Carlo method (RxMC)\cite{smith94c, johnson94a}.
In brief, this method uses the Metropolis-Hastings algorithm\cite{metropolis53a, hastings70a} to sample from a semi-grandcanonical distribution under the constraints enforced by the stoichiometry of the reactions.  
For each reaction step, one of the reactions and its direction is selected randomly with a uniform probability.
Next, according to the stoichiometry, particles are added at a random position to the simulation box and/or randomly selected and deleted or have their identity changed.  
This proposed new configuration (n) is then accepted according to the criterion
\begin{align}
\begin{split}
P^{\mathrm{RxMC}}_{\textrm{n,o}} =& \min \Biggl\{ 1,
	\left( \prod_{i} \frac{N_i^{\textrm{o}}! \left(V \cref\right)^{\nu_i \xi}}{(N_i^{\textrm{o}}+\nu_i \xi)!}\right)\times\\
&\times\exp\left(\beta \left[ \xi \sum_i \nu_i (\mu_i - \muref_i) - \Delta \mathcal{U}_{\textrm{n,o}}\right] \right) \Biggr\},
\label{eq:sim:g-rxmc}
\end{split}
\end{align}
where $\beta$ is the inverse temperature, $V$ the simulation box volume, $\cref=1\,$M is the reference concentration, $\mu_i$ is the chemical potential of species $i$, $\muref_i$ is the reference chemical potential of species $i$, $\Delta \mathcal{U}_{\textrm{n,o}}=\mathcal{U}_{\textrm{n}}-\mathcal{U}_{\textrm{o}}$ is the change in potential energy, $\nu_i$ is the stoichiometric coefficient of species $i$ and $\xi$ is the extent of reaction which takes the value $\xi=1$ for the forward and $\xi=-1$ for the reverse reaction.
If the new configuration is rejected, the previous configuration (o) is kept.
From \autoref{eq:sim:g-rxmc} it is obvious that the method takes as its input a set of chemical potentials $\mu_i - \muref_i$ or equivalently a set of equilibrium constants.
Here an important question arises: How should one choose these input parameters in order to achieve a desired reservoir composition? This will be answered in the following section.
In addition to sampling the different chemical compositions of the system, one must also sample different conformations of the polymer chains and particles. 
To do this one can use either MC techniques or molecular dynamics.
In the following, we make use of both Metropolis-MC and  Langevin MD.\cite{langevin08a, allen87a} 

\section{Determining the Reaction Constants}
\label{sec:det_re_c}
For the kind of setup considered here, the reservoir is fully characterized by three parameters, for instance $K_{\ce{H+},\ce{Cl-}}$, $K_{\ce{Na+},\ce{Cl-}}$ and $K_{\ce{H$_{n}$a}}$.
(There are of course also the $n$ p$K_\mathrm{a}$-values p$K_\mathrm{a}^1$, p$K_\mathrm{a}^2$, ..., p$K_\mathrm{a}^n$, however these are simply input parameters specific to the simulated acid.)
In order to mimic experiments, we typically want to impose on the reservoir the pH-value, $\mathrm{pH}^{\mathrm{res}}$, the salt concentration, $c_{\mathrm{NaCl}}^\mathrm{res}$, and the total concentration, $c_{\ce{H_$n$ a}}^{\mathrm{res},0}$.
Because the input parameters for the simulations are the equilibrium constants (or equivalently the chemical potentials), finding the correct equilibrium constants to achieve a desired reservoir composition amounts to a so-called ``inverse problem'': we need to find the correct ``cause'' (the equilibrium constants) to achieve a desired ``effect'' (the reservoir composition).
Landsgesell et al.\cite{landsgesell20b} described  two distinct ways to determine the required reactions constants, both of which rely on auxiliary simulations of the reservoir:

\begin{enumerate}
 \item \textbf{Approach using Widom Particle Insertion}: In order to determine the reactions constants, we need to know the relationship between the activity coefficients and the concentrations of the various $z$-valent ions.
 The simplest way to determine this relation is to simulate a sufficiently large box of small ions at a range of different concentrations of the various ion types and to determine the excess chemical potential for different ion pairs, triplets, etc. using the method of Widom particle insertion.\cite{widom63a} (Of course one can in principle also use a semi-empirical formula like the Davies equation,\cite{atkins10a} however the range of applicability of such an approach is inherently limited.)
 The $n+1$ resulting chemical potentials are thus $n+1$-dimensional functions on a grid that can then be interpolated. 
 In combination with the definition of the pH, the law of mass-action of the autoionization of water and the law of mass-action for the $n$ various acid-dissociation reactions of \ce{H_$n$ a} and its ionized forms, this results in a set of nonlinear equations that can be solved in a self-consistent loop to ultimately yield the desired equilibrium constants.
 While this approach is in principle as exact as desired, it quickly becomes unfeasible as $n$ grows, since the number of auxiliary simulations that are required grows exponentially with $n$.
 \item \textbf{Calibration Method}: Alternatively, one may simply impose values for the equilibrium constants (for instance using the ideal gas, Debye-Hückel theory or a semi-empirical formula as a starting point) and then run a reservoir simulation with these values. 
 Afterwards one can calculate the reservoir composition from the simulation and slightly adjust the equilibrium constants. 
 This results in an iterative procedure that stops once the desired accuracy is achieved.
 Since this approach requires in general multiple simulations and manual adjustments to achieve a desired reservoir composition, it can become cumbersome or unfeasibly long.
\end{enumerate}
Because both of these methods have their difficulties, we here propose an alternative approach, that can be viewed as a more sophisticated version of the calibration method.
Our new approach generalizes a recently developed method to dynamically tune the chemical potential in a \emph{single} grand-canonical simulation to achieve a desired particle number.\cite{miles22a}
For the reader unfamiliar with the original method, we shortly recap the essential points before describing how it can be applied to the current system.
For convenience, we adopt the notation of Ref.\cite{miles22a}
The method makes use of the fact that the derivative of the chemical potential $\mu$ with respect to the particle number $N$ (``compressibility'' $\kappa$) can be expressed in terms of the variance $\mathrm{Var}\lbrack N\rbrack$ of the particle number $N$:
\begin{align}
\kappa = \frac{\mathrm{d}\left\langle N\right\rangle}{\mathrm{d}\mu} = \beta\,\mathrm{Var}\lbrack N\rbrack.
\label{eq:grand-canonical-compressibility}
\end{align}
Using $\kappa$, in a grand-canonical simulation with desired particle number $N^*$ one can make an initial guess $\mu_{t=0}$ for the chemical potential and then periodically update $\mu_t$ after a certain number of Monte-Carlo steps according to the linearized formula
\begin{align}
\mu_{t+1} = \overline{\mu}_t + \frac{N^*-\overline{N}_t}{\overline{\kappa}_t},
\label{eq:update_scheme_mu}
\end{align}
where $\overline{x}_t$=$\overline{x}^t$ denotes a time average of the quantity $x$.
In the following, we will use the average over the more recent half of the trajectory:
\begin{align}
\overline{x}_t  = \frac{1}{L_t}\sum_{t'=\lceil t/2\rceil}^{t}x_{t'}.
\end{align}
Simply using \autoref{eq:grand-canonical-compressibility} to calculate $\overline{\kappa}_t$ results in the fluctuation-based estimator 
\begin{align}
 \kappa_t^\mathrm{fluc} = \beta\,\overline{\mathrm{Var}}_t\lbrack N\rbrack
\end{align}
that is calculated as the variance of $N$ over the more recent half of the trajectory and which is only useful at later times. 
To guarantee reasonable values of $\overline{\kappa}_t$ also at early times, one uses
\begin{align}
 \overline{\kappa}_t = \max\left\lbrack \kappa_t^\mathrm{min},\min\left(\kappa_t^\mathrm{max},\kappa_t^\mathrm{fluc}\right)\right\rbrack 
 \label{eq:kappa_estimator}
\end{align}
with the bounds
\begin{align}
\kappa_t^\mathrm{min} &= \frac{\alpha}{\sqrt{t+1}}\\
\kappa_t^\mathrm{max} &= \sqrt{\frac{\overline{\mathrm{Var}}_t\lbrack N\rbrack}{\overline{\mathrm{Var}}_t\lbrack \mu\rbrack}},
\end{align}
where $\alpha\propto V/U$ with the system volume $V$ and a characteristic energy scale $U$.
\autoref{eq:update_scheme_mu} in combination with \autoref{eq:kappa_estimator} results in a robust update scheme for $\mu_t$ that eventually converges to the correct value.
To avoid fully recalculating $\overline{x}_t$ and $\overline{\mathrm{Var}}_t\lbrack x\rbrack$ at every time step, they can be updated incrementally using a modified Welford algorithm.\cite{welford62a, miles22a}

For the kinds of systems under consideration here, we need to tune two chemical potentials: $\mu_{\ce{NaCl}}$ to achieve a desired salt concentration, $c_{\mathrm{NaCl}}^\mathrm{res}$, and $\mu_{\ce{H_$n$ a}}$ to achieve a desired total concentration of the dissolved acid, $c_{\ce{H_$n$ a}}^{\mathrm{res},0}$.
Because the $n$ p$K_\mathrm{a}$-values p$K_\mathrm{a}^1$, p$K_\mathrm{a}^2$, ..., p$K_\mathrm{a}^n$ are fixed, the $n$ chemical potentials $\mu_{\ce{H$_{n-z}$ a$^{z-}$}}$ for $z=1,...,n$ are completely determined.
In general, to apply \autoref{eq:update_scheme_mu} to a multi-component system, $\kappa$ needs to be promoted to a matrix $\kappa_{ij}$.
However, for the present application it turns out that neglecting the off-diagonal elements of $\kappa_{ij}$ and applying the tuning procedure to each species independently is sufficient.
Since the results can always be checked for consistency, this approximation is inherently safe.
As we show below, the procedure reliably converges for a wide range of system parameters.

For the salt $\ce{NaCl}$ we simply apply the original $\mu$-tuning method as described above. 
The instantaneous number of $\ce{NaCl}$ ion pairs can be measured as
\begin{align}
 N_{\ce{NaCl}}^t = \min\left(N_{\ce{Na+}}^t,N_{\ce{Cl-}}^t\right).
\end{align}
Note that in general we have either $N_{\ce{Na+}}>N_{\ce{NaCl}}$ or $N_{\ce{Cl-}}>N_{\ce{NaCl}}$, because one needs to add \ce{NaOH} or \ce{HCl} to adjust the pH to the desired value.
Applying the method thus ultimately results in a dynamically evolving equilibrium constant $K_{\ce{Na+},\ce{Cl-}}^t=\exp(\beta\mu_{\ce{NaCl}}^t)$.
For the dissolved acid, we make use of the identity (see appendix for a detailed calculation)
\begin{align}
\begin{split}
\tilde{\kappa}^\mathrm{fluc} &= \frac{\partial}{\partial\mu_{\ce{H_$n$ a}}}\left\langle N_{\ce{H_$n$ a}}^0\right\rangle = \sum_{z=0}^{n} \frac{\partial}{\partial\mu_{\ce{H$_n$ a}}} \left\langle N_{\ce{H$_{n-z}$ a$^{z-}$}}\right\rangle\\
&= \beta\sum_{z=0}^{n} \mathrm{Cov}\lbrack N_{\ce{H$_{n-z}$ a$^{z-}$}}, N_{\ce{H_$n$ a}}\rbrack,
\end{split}
\end{align}
where $\mathrm{Cov}\lbrack x,y\rbrack$ denotes the covariance of $x$ and $y$,
to arrive at the update rule 
\begin{align}
\mu_{\ce{H_$n$ a}}^{t+1} = \overline{\mu_{\ce{H_$n$ a}}}^{t} + \frac{N_{\ce{H_$n$ a}}^0-\sum_{z=0}^{n}\overline{N_{\ce{H$_{n-z}$ a$^{z-}$}}}^t}{\overline{\tilde{\kappa}}_t}.
\end{align}
Here, $\overline{\tilde{\kappa}}_t$ is defined by 
\begin{align}
 \overline{\tilde{\kappa}}_t = \max\left\lbrack \tilde{\kappa}_t^\mathrm{min},\min\left(\tilde{\kappa}_t^\mathrm{max},\tilde{\kappa}_t^\mathrm{fluc}\right)\right\rbrack 
\end{align}
with
\begin{align}
\tilde{\kappa}^\mathrm{fluc}_t =& \sum_{z=0}^{n} \overline{\mathrm{Cov}}_t\lbrack N_{\ce{H$_{n-z}$ a$^{z-}$}}, N_{\ce{H_$n$ a}}\rbrack\\
\tilde{\kappa}_t^\mathrm{min} =& \frac{\alpha}{\sqrt{t+1}}\\
\begin{split}
\tilde{\kappa}_t^\mathrm{max} =& \mathrm{sgn}\left(\sum_{z=0}^{n}\overline{\mathrm{Cov}}_t\lbrack N_{\ce{H$_{n-z}$ a$^{z-}$}}, N_{\ce{H_$n$ a}}\rbrack\right)\times\\
&\times\sqrt{\frac{\left|\sum_{z=0}^{n}\overline{\mathrm{Cov}}_t\lbrack N_{\ce{H$_{n-z}$ a$^{z-}$}}, N_{\ce{H_$n$ a}}\rbrack\right|}{\overline{\mathrm{Var}}_t\lbrack \mu\rbrack}},
\end{split}
\end{align}
where we have again $\alpha\propto V/U$.
As we show in the appendix, there exists a simple formula to update $\overline{\mathrm{Cov}}_t\lbrack x,y\rbrack$ incrementally.
Analogous to the previous case, applying the method results in a dynamically evolving equilibrium constant $K_{\ce{H_$n$ a}}^t$.

In our reservoir simulations, we tune $\mu_{\ce{NaCl}}$ and $\mu_{\ce{H_$n$ a}}$ simultaneously, i.e. we periodically update both $K_{\ce{Na+},\ce{Cl-}}^t$ and $K_{\ce{H_$n$ a}}^t$ after a certain number of Monte-Carlo steps.
As a consequence, the equilibrium constants for the insertion moves, $K_{(z-l)\ce{H+},\,l\ce{Na+},\,\ce{H$_{n-z}$a$^{z-}$}}^t$, also become time-dependent and are updated in each loop as well.
There is still a subtlety concerning the equilibrium constant $K_{\ce{H+},\ce{Cl-}}^t$ (and thus also $K_{\ce{Na+},\ce{OH-}}^t$), which is time-dependent as well: In the special case $n=1$ one can use the charge neutrality condition
\begin{align}
c_{\ce{H+}} + c_{\ce{Na+}} = c_{\ce{OH-}} + c_{\ce{Cl-}} + c_{\ce{a-}}
\end{align}
and multiply by the mean activity coefficient $\sqrt{\gamma} = \sqrt{\gamma_+\gamma_-}$
to get 
\begin{align}
\begin{split}
c_{\ce{Na+}}\sqrt{\gamma} =& c_{\ce{OH-}}\sqrt{\gamma} + c_{\ce{Cl-}}\sqrt{\gamma} - c_{\ce{H+}}\sqrt{\gamma} + c_{\ce{a-}}\sqrt{\gamma}.
\end{split}
\end{align}
Using the law of mass-action and the definition of the pH, we arrive at
\begin{align}
\begin{split}
c_{\ce{Na+}}\sqrt{\gamma} =& \cref 10^{-\left(\mathrm{pH}-14\right)} + c_{\ce{Cl-}}\sqrt{\gamma} - \cref 10^{-\mathrm{pH}}\\
&+ \frac{K_{\mathrm{a}}K_{\mathrm{Ha}}\cref}{10^{-\mathrm{pH}}}.
\end{split}
\end{align}
Inserting this expression into the definition of $K_{\ce{Na+},\ce{Cl-}}$ results in a quadratic equation 
\begin{align}
\begin{split}
 K_{\ce{Na+},\ce{Cl-}}(\cref)^2 =& \left(c_{\ce{Cl-}}\sqrt{\gamma}\right)^2 + c_{\ce{Cl-}}\sqrt{\gamma}\cref 10^{-\left(\mathrm{pH}-14\right)}\\
 &-c_{\ce{Cl-}}\sqrt{\gamma}\left(\cref 10^{-\mathrm{pH}} - \frac{K_{\mathrm{a}}K_{\mathrm{Ha}}\cref}{10^{-\mathrm{pH}}}\right)
\end{split}
\end{align}
which can be solved for $c_{\ce{Cl-}}\sqrt{\gamma}$:
\begin{align}
\begin{split}
 c_{\ce{Cl-}}\sqrt{\gamma} =&-\frac{1}{2}\left(\cref 10^{-\left(\mathrm{pH}-14\right)} - \cref 10^{-\mathrm{pH}} + \frac{K_{\mathrm{a}}K_{\mathrm{Ha}}\cref}{10^{-\mathrm{pH}}}\right)\\ 
 &+ \frac{1}{2}\Biggl(\left(\cref 10^{-\left(\mathrm{pH}-14\right)} - \cref 10^{-\mathrm{pH}} + \frac{K_{\mathrm{a}}K_{\mathrm{Ha}}\cref}{10^{-\mathrm{pH}}}\right)^2\\
 &+4K_{\ce{Na+},\ce{Cl-}}(\cref)^2\Biggr)^{\frac{1}{2}}
\end{split}
\end{align}
This thus enables one to give the exact value of $K_{\ce{H+},\ce{Cl-}}$:
\begin{align}
 K_{\ce{H+},\ce{Cl-}}= \frac{c_{\ce{H+}}c_{\ce{Cl-}}}{(\cref)^2}\gamma = 10^{-\mathrm{pH}} \frac{c_{\ce{Cl-}}}{\cref}\sqrt{\gamma}
 \label{eq:khcl}
\end{align}
For $n>1$ this is not possible due to complications that arise because of the multivalent terms.
An approach that works in this case is as follows:
$K_{\ce{H+},\ce{Cl-}}$ is given by \autoref{eq:khcl}, where $\gamma$ is the activity coefficient of a monovalent ion pair. 
To approximately calculate $\gamma$ in the simulation, we can use
\begin{align}
\gamma^t = \frac{K_{\ce{Na+},\ce{Cl-}}^t (\cref)^2}{\overline{c_{\ce{Na+}}}^t\overline{c_{\ce{Cl-}}}^t}
\label{eq:approximate_activity}
\end{align}
and thus get
\begin{align}
K_{\ce{H+},\ce{Cl-}}^t = 10^{-\mathrm{pH}}\sqrt{\frac{\overline{c_{\ce{Cl-}}}^t}{\overline{c_{\ce{Na+}}}^t}K_{\ce{Na+},\ce{Cl-}}^t}.
\label{eq:approximate_khcl}
\end{align}
While this expression is not exact, it converges to the correct value for large times as we show in the appendix.
In the following, we always use the approximate scheme, if not stated otherwise.

\section{Numerical Test: Determining the Reaction Constants}
\label{sec:mu-tuning}

\subsection{Setup}
To showcase our method, we ran a number of test simulations for systems which can exchange a small acid with a reservoir.
In a first step, we show the validity of the dynamical tuning algorithm for the chemical potential to arrive at the desired reservoir composition.
Here, we focus on the case of a monoprotic acid. 
The analogous results for a diprotic acid are reported in the appendix.

For all of the tests described below we used the following simulation setup and protocol:
We carried out bulk simulations in a box with periodic boundary conditions.
The size of the cubic simulation box is set in such a way that according to an ideal gas estimate the number of particles of the most numerous species is $N=500$.
We carry out tests both for an ideal system and an interacting system.
In the interacting case, as our simulation model we employ the restricted primitive model (RPM).\cite{hansen13a}
In the RPM, ions are represented by spheres with an excluded volume interaction, in our case modeled by a WCA potential\cite{weeks71a}
\begin{align}
V_\text{WCA}(r) = \begin{cases}4\epsilon\left(\left(\frac{\sigma}{r}\right)^{12}-\left(\frac{\sigma}{r}\right)^6\right)+\epsilon &\mbox{if } r\leq2^\frac{1}{6}\sigma\\0 &\mbox{if }r> 2^\frac{1}{6}\sigma\end{cases}
\label{eq:wca}
\end{align}
with diameter $\sigma=0.355\,\text{nm}$ and an energy of $\epsilon=k_\text{B}T$.
These explicit ions are embedded into an implicit solvent and also interact via the Coulomb potential
\begin{align}
V^{ij}_\text{Coulomb}(r) = \frac{\lambda_\text{B}k_\text{B}Tz_iz_j}{r}
\label{eq:Coulomb-Potential}
\end{align}
where $z_i$ is the valency of species $i$ and the Bjerrum length $\lambda_\text{B}=e^2/4\pi\epsilon k_\text{B}T$ is set to a value of $\lambda_\text{B}=2\sigma=7.1\,\text{\r{A}}$ which accounts for the dielectric properties of water at room temperature ($T\approx 300\,\text{K}$).
In our simulations, we use the P$^3$M algorithm\cite{hockney88a} with a relative error\cite{deserno98a, deserno98b} of $10^{-3}$ to sum up the electrostatic energies and forces. 

For the ideal system, we start the simulations at random initial particle numbers.
In the interacting case, we use an extended Debye-Hückel formula (Davies-equation) estimate as our starting point.
In both cases, we perform a total of $10^6$ loops which consist of 10 reaction steps each.
For the interacting system, we also include 10 MC single-particle displacement moves in each loop.
These help to decorrelate the system faster than mere reaction moves.
After each loop, we update the chemical potentials and thus the equilibrium constants according to the method introduced above.
We set $\alpha = 0.1$.
We ran tests for p$K_{\mathrm{a}} = 4.0$, $c_{\mathrm{NaCl}}^\mathrm{res}\in\left\lbrace 0.1\,\mathrm{M}, 0.03\,\mathrm{M}, 0.01\,\mathrm{M}\right\rbrace$, $c_{\ce{Ha}}^{\mathrm{res},0}\in\left\lbrace 0.1\,\mathrm{M}, 0.03\,\mathrm{M}, 0.01\,\mathrm{M}\right\rbrace$ and $\mathrm{pH}^\mathrm{res}\in\left[1.0,13.0\right]$.
A real acid with p$K_{\mathrm{a}} \approx 4.0$ is for instance acrylic acid.
All simulations were carried out using the simulation software package ESPResSo.\cite{weik19a}

\subsection{Ideal System}
The ideal limit provides an important test case, as it is accessible to an analytical solution (details are reported in the appendix).
In \autoref{fig:main_figure_2} (a) we show the evolution of the mean concentrations $\overline{c_{\mathrm{NaCl}}}^t$ and $\overline{c_{\ce{Ha}}^{0}}^{t}$ for all simulated parameter combinations.
To make the different simulations comparable, all concentrations are normalized by the respective desired concentration.
As the plot shows, the mean concentrations converge to the desired values for all simulated parameters.

Because in the ideal case we can calculate the concentrations of all species analytically, we also compare these to the simulation results. 
The plot in \autoref{fig:main_figure_2} (b) shows the mean concentration $\overline{c_i}^t$ at the end of the simulation for the different species as a function of the respective ideal result. 
As the plot shows, all data points collapse (almost) perfectly onto the bisecting line.
To suppress finite-size effects, only data points with a mean number of particles larger than 10 were included.
We include an analogous comparison with the ideal theory for the equilibrium constants in the appendix (\autoref{fig:esi_figure_1}), where we observe again a very good agreement.

\begin{figure}[H]
\centering
\includegraphics[width=\columnwidth]{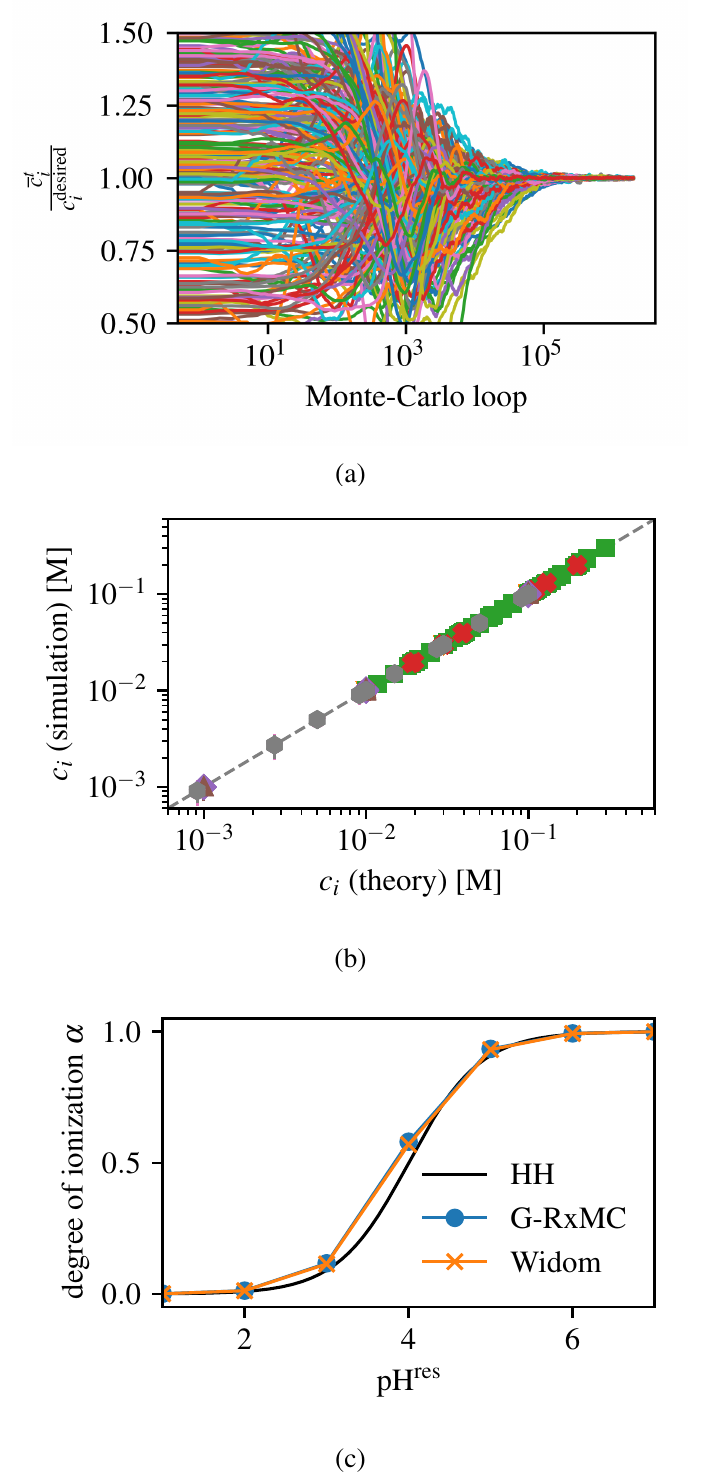}
\caption{\label{fig:main_figure_2}(a): Evolution of the mean concentrations $\overline{c_{\mathrm{NaCl}}}^t$ and $\overline{c_{\ce{Ha}}^{0}}^{t}$ (normalized by the resepective desired concentration) for an ideal system for all simulated parameter combinations.
(b): Mean concentrations of the various species measured from the simulation for an ideal system (at the very end) vs. the concentrations predicted by the ideal theory.
(c): Plot of the degree of ionization of the small acid particles for an interacting system as a function of pH for $c_{\ce{NaCl}}^\text{res}=0.1$\,M and $c_{\ce{Ha}}^\text{res,0}=0.1$\,M. The plot compares the ideal result given by the Henderson-Hasselbalch equation (HH) with the result from the generalized G-RxMC method and the result obtained from a semi-analytical calculation using data from Widom insertion simulations.}
\end{figure}

\subsection{Interacting System}
We perform analogous tests for an interacting system with the interactions specified above.
As a test, we compare the simulation results for this interacting system with a calculation that uses data from Widom insertion\cite{widom63a} simulations for a monovalent salt solution (compare the appendix for an explanation).
In \autoref{fig:main_figure_2} (c), we show the resulting ionization curve obtained using this procedure for the most non-ideal case considered here ($c_{\ce{NaCl}}^\text{res}=0.1$\,M and $c_{\ce{Ha}}^\text{res,0}=0.1$\,M.), which is in excellent agreement with the simulation results, while there are deviations from the ideal prediction. 
The enhanced ionization as compared to the ideal result is expected, as the excess chemical potential of a salt solution is negative in the range of concentrations considered here.
Some more tests for correctness and internal consistency of the results are included in the appendix.

\section{An Example System: Simulation of a Weak Polybase Solution Coupled to a Reservoir Containing a Weak Diprotic Acid}

\subsection{Setup}
Having shown that the propsed tuning procedure for the reservoir composition works correctly, we can now apply the method to a simple test system.
Here we consider a solution of polybase molecules which is coupled to a reservoir at a given pH, salt concentration and concentration of a weak diprotic acid.
All particles except the polybase chains can be exchanged between the two phases.
Experimentally, such a setup could be realized by coupling the polybase solution to an aqueous solution via a semi-permeable membrane which does not allow the polybase chains (but all other particles) to pass.
In the following, we denote the conjugate acid of the basic monomers by \ce{BH+} (i.e. the ionized/protonated state) and the neutral monomers by \ce{B}. 
This implies that we can write the reaction in the form 
\begin{align}
 \ce{BH+ <=> B + H+}
\end{align}
with an equilibrium constant which we set to p$K_\text{A}^\text{base}=10.0$ in the following.
Although we do not make an attempt to exactly match an experimentally realized system, this could for instance be regarded as a simple model for a polypeptide chain consisting of Lysine (p$K_\text{A}^\text{base}\approx10.4$).

We retain the RPM for ions described above. 
To model the polybase chains, we combine the RPM with a generic bead-spring model for polymers derived from the well-established Kremer-Grest model.\cite{grest86a}
This kind of model, which has been applied to a wide range of polyelectrolyte systems in the past,\cite{yin08a, stano20b, landsgesell22a, beyer22a} accurately models the electrostatic and steric interactions while neglecting more chemically specific effects.
In more detail, analogous to the RPM, all monomers (both neutral and charged) interact via a soft-sphere WCA interaction as described by \autoref{eq:wca}.
Furthermore, charged monomers interact via the Coulomb potential given by \autoref{eq:Coulomb-Potential}.
The covalent bonds between monomers are modelled by nonlinear springs, described by the FENE (Finite Extensibility Nonlinear Elastic) potential
\begin{align}
  V_\text{FENE}(r) = \begin{cases}-\frac{k\Delta r_\text{max}^2}{2}\ln\left(1-\left(\frac{r-r_0}{\Delta r_\text{max}}\right)^2\right) &\mbox{if } r\leq \Delta r_\text{max}\\ \infty &\mbox{if }r>\Delta r_\text{max}.
  \end{cases}
\end{align}
For the present study, we set the maximum bond extension to $\Delta r_\text{max} = 1.5\sigma$, the spring constant to $k=30k_\text{B}T/\sigma^2$ and the equilibrium bond extension to $r_0=0$.
We set the p$K_\text{a}$-values of the weak diprotic acid to p$K_\text{a}^1=4.0$ and p$K_\text{a}^2=7.0$.
While we do not mimic any specific system, these values are completely in the range that could be realized experimentally (e.g. malonic acid with p$K_\text{a}^1\approx 2.8$ and p$K_\text{a}^2\approx5.7$).
We ran simulations for a total of 16 chains of length $N=50$ at $c_{\mathrm{NaCl}}^\mathrm{res}= 0.1\,\mathrm{M}$, $c_{\ce{H2a}}^{\mathrm{res},0}=0.1\,\mathrm{M}$, $\mathrm{pH}^\mathrm{res}\in\left[1.0,13.0\right]$ and different monomer concentrations $c_{\ce{mon}}\in\left\lbrace 0.1\,\mathrm{M}, 0.03\,\mathrm{M}, 0.01\,\mathrm{M}\right\rbrace$.
In effect, increasing the concentration of monomers inside the system increases the Donnan effect.

\begin{figure}[H]
\centering
\includegraphics[width=\columnwidth]{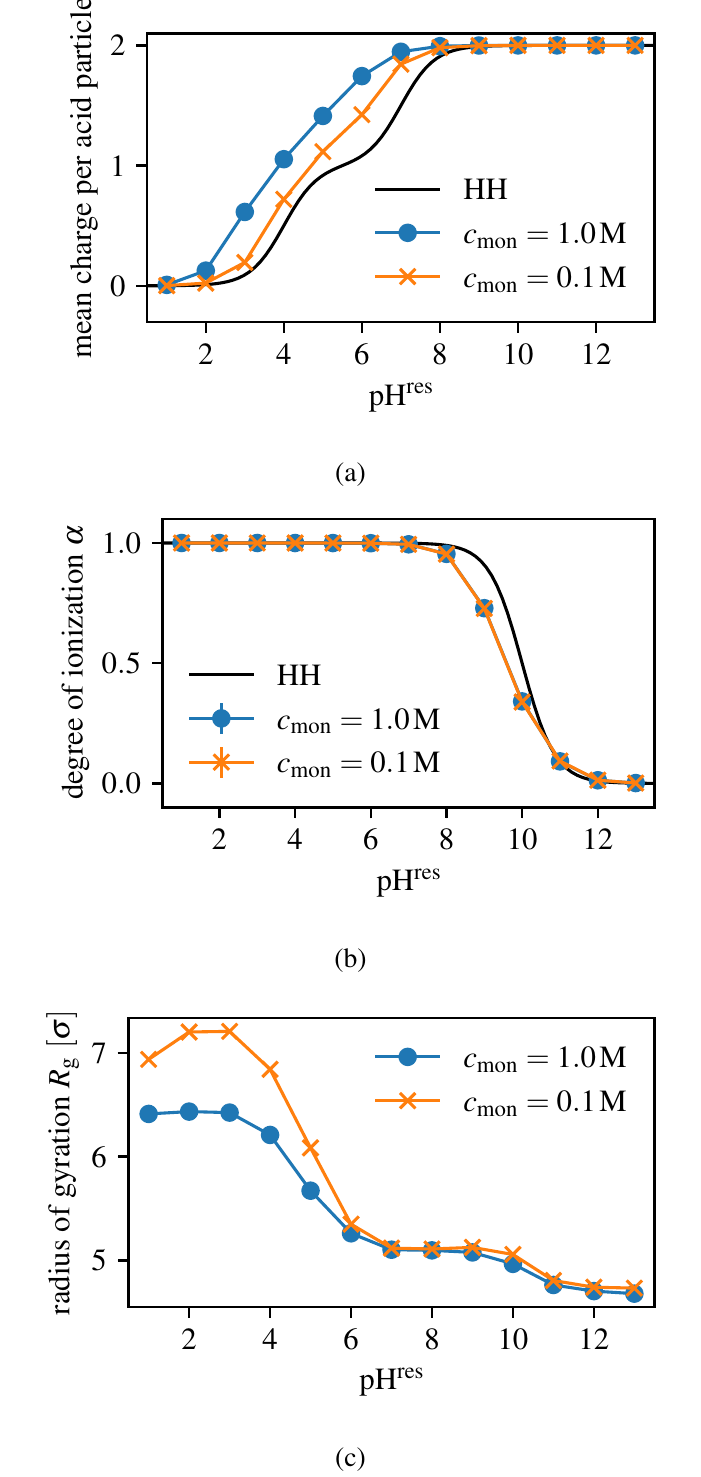}
\caption{\label{fig:main_figure_3}
Various plots for the interacting test system.
The shown plots correspond to $c_{\ce{NaCl}}^\text{res}=0.1$\,M and $c_{\ce{H2a}}^\text{res,0}=0.1$\,M and different monomer concentrations $c_{\ce{mon}}$.
(a): Plot of the absolute value of the mean charge per diprotic acid particle vs the pH in the reservoir.  As a comparison, the prediction by the ideal theory (Henderson-Hasselbalch equation) is also shown.
(b): Plot of the degree of ionization of the weak polybase as a function of pH$^\mathrm{res}$.
  The prediction according to the ideal theory (Henderson-Hasselbalch equation) is shown as well.
(c): Plot of the mean radius of gyration of the chains as a function of pH$^\mathrm{res}$.
}
\end{figure}

\subsection{Results}
As before, we first carried out tests for an ideal system which can be compared to the analytical solution.
These results can be found in the appendix.
For the interacting system, we observe strong deviations from the ideal prediction for the mean charge per acid particle, as shown in \autoref{fig:main_figure_3} (a).
In essence, the charge of the diprotic acid is enhanced as compared to the ideal prediction. 
Part of this enhancement is caused by the Donnan effect, which effectively increases the pH inside the system as compared to pH$^\text{res}$. 
Furthermore, the ionization is also increased by the electrostatic interactions, which lead to a negative excess chemical potential of the ionized species. 
The increase of the shift we observe with increasing the monomer concentration in the system is thus also of two-fold origin: On the one hand, an increasing monomer concentration leads to a stronger Donnan effect. 
On the other hand, the increasing concentration of the monomers (which are themselves positively charged in the considered pH-regime) further lowers the excess chemical potential of the ions, leading to an increased ionization.

For the weak polybase, we observe an ionization behaviour that is (as expected) suppressed as compared to the Henderson-Hasselbalch equation (\autoref{fig:main_figure_3} (b)).
As in the case of the weak diprotic acid, the shift is caused by a combination of the Donnan effect and the electrostatic interactions. 
Interestingly, the observed shift is only minor as compared to the rather large shifts typically observed in simulations of weak polyacids or -bases in monovalent salt solutions.\cite{landsgesell19a}
This effect is a consequence of a partial cancellation of the strong repulsion of the ionized basic monomers and the strong attraction of the monomers and the divalent ions.\cite{stano20b}

Finally, in \autoref{fig:main_figure_3} (c) we show the mean radius of gyration $R_\text{g}$ of the weak polybase chains as a function of pH$^\mathrm{res}$.
The swelling behaviour of the chains is much more complex than for instance in the simpler case of chains which are coupled to a reservoir containing a strong, monovalent salt.\cite{landsgesell20b}
The most obvious feature of the swelling curves is that the swelling proceeds in a two-step manner when going from a high value of pH$^\mathrm{res}$ to a low value. 
This can be explained in the following way: at very high values of pH$^\mathrm{res}$ (pH$^\mathrm{res}>11$) the polybase chains are essentially neutral and thus attain a fairly compact conformation. 
Once the pH-value is lowered, the chains become ionized and begin to swell.
However, this swelling is only weak, because the chains stay fairly collapsed due to the presence of the doubly ionized acid particles, which act as divalent counterions.
Interestingly, at this the swelling is almost the same for both investigated values of the monomer concentration.
This behaviour is in agreement with the previous observation that beyond a certain threshold the addition of more multivalent counterions does not influence the conformational behaviour of a weak polyelectrolyte anymore.\cite{stano20b}
The second swelling step, beginning around pH$^\mathrm{res}\approx6$ is triggered by the doubly ionized weak diprotic acid particles first becoming monovalent and finally neutral. 
Here we observe that the swelling is more pronounced at a lower monomer concentration. 
This is because the Donnan effect is smaller in this case and thus the ionic strength inside the system is lower, leading to less screening of the electrostatic interactions.
Finally, at very low values of pH$^\mathrm{res}$, the swelling decreases again. 
This efffect, similar to what is for instance also observed in weak polyelectrolyte hydrogels,\cite{landsgesell22a} is caused by the increasing ionic strength in the system due to the addition of \ce{HCl}, which increases the screening between the ionized base monomers.

\section{Summary and Outlook}
To summarize, we introduced a generalized grand-reaction method to
model the exchange of weak (polyprotic) acids and bases between a
reservoir and a polyelectrolyte phase.  To the best of our knowledge
for the first time this new method makes it possible to investigate
the partitioning of weak polyprotic acids between a solution and a
polymeric phase such as a hydrogel.  Because the resulting reservoir
is now itself a much more complex system, the existing approaches to
extract the required equilibrium constants that correspond to a
desired reservoir composition are not feasible anymore.  In order to
solve this problem, we generalized the $\mu$-tuning algorithm by Miles
et al.\cite{miles22a}.

We performed extensive numerical tests in order to validate our
proposed tuning method for the chemical potentials.  Finally, we
investigated a simple test system consisting of a weak polybase
coupled to a solution containing a weak diprotic acid.  As a
consequence of the interplay of ion partioning and charge regulation
effects (both of the weak polybase and the weak diprotic acid) we
observed a two-step swelling behaviour of the chains.

Through our generalization, the method can now be applied to study the
partitioning of weak polyprotic acid molecules between a solution and
a weak polyelectrolyte hydrogel.  The study of the same effects in
related systems such as coarcervates would also be feasible.
Furthermore, it should in principle also be possible to apply the
described method to the partitioning of more complex particles such as
short polypeptides.  In this case, however, one would need to combine
it with a biased Monte-Carlo scheme in order to ensure a high enough
acceptance probability for insertion moves.\cite{smit95a}

\begin{acknowledgments}
\noindent CH acknowledges funds by the German Research Foundation (DFG) -- grants No.\ 451980436 
and No.\ 268449726. 
Parts of this work were also performed within the collaborative framework of the research unit \textit{Adaptive Polymer Gels with Controlled Network Structure (FOR2811)}, funded by the German Research Foundation under No. 397384169.
DB acknowledges helpful discussions with Mariano Brito and Jonas Landsgesell.
\end{acknowledgments}

\section*{Data Availability Statement}
\noindent The simulation scripts and data that support the findings of this study are available upon reasonable request.

\section*{Conflict of Interest}

\noindent The authors have no conflicts to disclose.
\text{ }\\
\appendix 

\section*{Author Contributions}
\noindent DB: Conceptualization, Formal Analysis, Methodology, Software, Writing (Original Draft)\\
\noindent CH: Conceptualization, Supervision, Writing (Review and Editing)

\section{Additional Calculations}
The identity 
\begin{align}
\frac{\partial}{\partial\mu_y} \langle N_x\rangle = \beta\mathrm{Cov}\lbrack N_x, N_y\rbrack
\end{align}
which is used in the main text, can be shown using a straightforward calculation in the grand-canonical ensemble.
As our starting point, we take the grand-canonical partition function for a system with $s$ different species (the same calculation also holds for a semi-grand-canonical ensemble), which can be expressed using the trace operator:
\begin{align}
\begin{split}
Z^\mathrm{G}\left(\left\{\mu_i\right\},V,T\right) &= \sum_{N_1=0}^{\infty}...\sum_{N_\mathrm{s}=0}^{\infty}Z\left(\left\{N_i\right\},V,T\right)\exp\left(\beta \sum_{i=1}^{N_\mathrm{s}}\mu_iN_i\right)\\
&=\mathrm{Tr}\left(\exp\left(-\beta\left(U- \sum_{i=1}^{N_\mathrm{s}}\mu_iN_i\right)\right)\right).
\end{split}
\end{align}
As is well known, the mean particle number can be expressed as
\begin{align}
\langle N_x\rangle = \frac{1}{\beta}\frac{\partial}{\partial\mu_x}\log Z^\mathrm{G} = \frac{1}{Z^\mathrm{G}}\mathrm{Tr}\left(N_x\exp\left(-\beta\left(U- \sum_{i=1}^{N_\mathrm{s}}\mu_iN_i\right)\right)\right).
\end{align}
Performing the derivative with respect to $\mu_y$, we arrive at
\begin{align}
\begin{split}
&\frac{1}{\beta}\frac{\partial}{\partial\mu_y} \langle N_x\rangle = \frac{1}{\beta}\frac{\partial}{\partial\mu_y} \frac{1}{Z^\mathrm{G}}\mathrm{Tr}\left(N_x\exp\left(-\beta\left(U- \sum_{i=1}^{N_\mathrm{s}}\mu_iN_i\right)\right)\right)\\
&= \frac{1}{Z^\mathrm{G}}\mathrm{Tr}\left(N_xN_y\exp\left(-\beta\left(U- \sum_{i=1}^{N_\mathrm{s}}\mu_iN_i\right)\right)\right)\\
&-\frac{1}{Z^\mathrm{G}}\mathrm{Tr}\left(N_x\exp\left(-\beta\left(U- \sum_{i=1}^{N_\mathrm{s}}\mu_iN_i\right)\right)\right)\times\\
&\times \frac{1}{Z^\mathrm{G}}\mathrm{Tr}\left(N_y\exp\left(-\beta\left(U- \sum_{i=1}^{N_\mathrm{s}}\mu_iN_i\right)\right)\right)\\
&= \langle N_x N_y\rangle - \langle N_x\rangle \langle N_y\rangle = \mathrm{Cov}\lbrack N_x, N_y\rbrack
\end{split}
\end{align}
which proves the identity.

\section{Additional Material for ``Numerical Test: Determining the Reaction Constants''}

In this first appendix, we include additional information and plots on the numerical tests we performed for the dynamical $\mu$-tuning procedure (\autoref{sec:mu-tuning}).

\subsection{Monoprotic Acid}

\subsubsection{Ideal System}
As mentioned in the main text, the ideal case is accessible to an analytical solution.
In particular, for a given pH$^\mathrm{res}$, salt concentration $c_{\mathrm{NaCl}}^\mathrm{res}$ and total concentration of the acid $c_{\ce{Ha}}^{\mathrm{res},0}$ one has the following concentrations:
\begin{align}
 c_{\ce{H+}}^{\mathrm{res}} &= \cref 10^{-\mathrm{pH}^\mathrm{res}}\\
 c_{\ce{OH-}}^{\mathrm{res}} &= \cref 10^{-\left(14-\mathrm{pH}^\mathrm{res}\right)}\\
 c_{\ce{Ha}}^{\mathrm{res}} &= \frac{c_{\ce{Ha}}^{\mathrm{res},0}}{1+\frac{\cref K_{\mathrm{a}}}{c_{\ce{H+}}^{\mathrm{res}}}}\\
 c_{\ce{a-}}^{\mathrm{res}} &= \frac{c_{\ce{Ha}}^{\mathrm{res},0}}{1+\frac{c_{\ce{H+}}^{\mathrm{res}}}{\cref K_{\mathrm{a}}}}\\
 c_{\ce{Na+}}^{\mathrm{res}} &= \max\left(c_{\ce{NaCl}}^{\mathrm{res}},c_{\ce{NaCl}}^{\mathrm{res}}+ c_{\ce{OH-}}^{\mathrm{res}}-c_{\ce{H+}}^{\mathrm{res}}+c_{\ce{a-}}^{\mathrm{res}}\right)\\
 c_{\ce{Cl-}}^{\mathrm{res}} &= \max\left(c_{\ce{NaCl}}^{\mathrm{res}},c_{\ce{NaCl}}^{\mathrm{res}} - c_{\ce{OH-}}^{\mathrm{res}}+c_{\ce{H+}}^{\mathrm{res}}-c_{\ce{a-}}^{\mathrm{res}}\right).
\end{align}
In addition, the equilibrium constants $K_{\ce{Na+},\ce{Cl-}}$, $K_{\ce{H+},\ce{Cl-}}$ and $K_{\ce{Ha}}$ can be simply calculated from these concentrations.

\begin{figure*}[htb]
\centering
\includegraphics[width=0.8\textwidth]{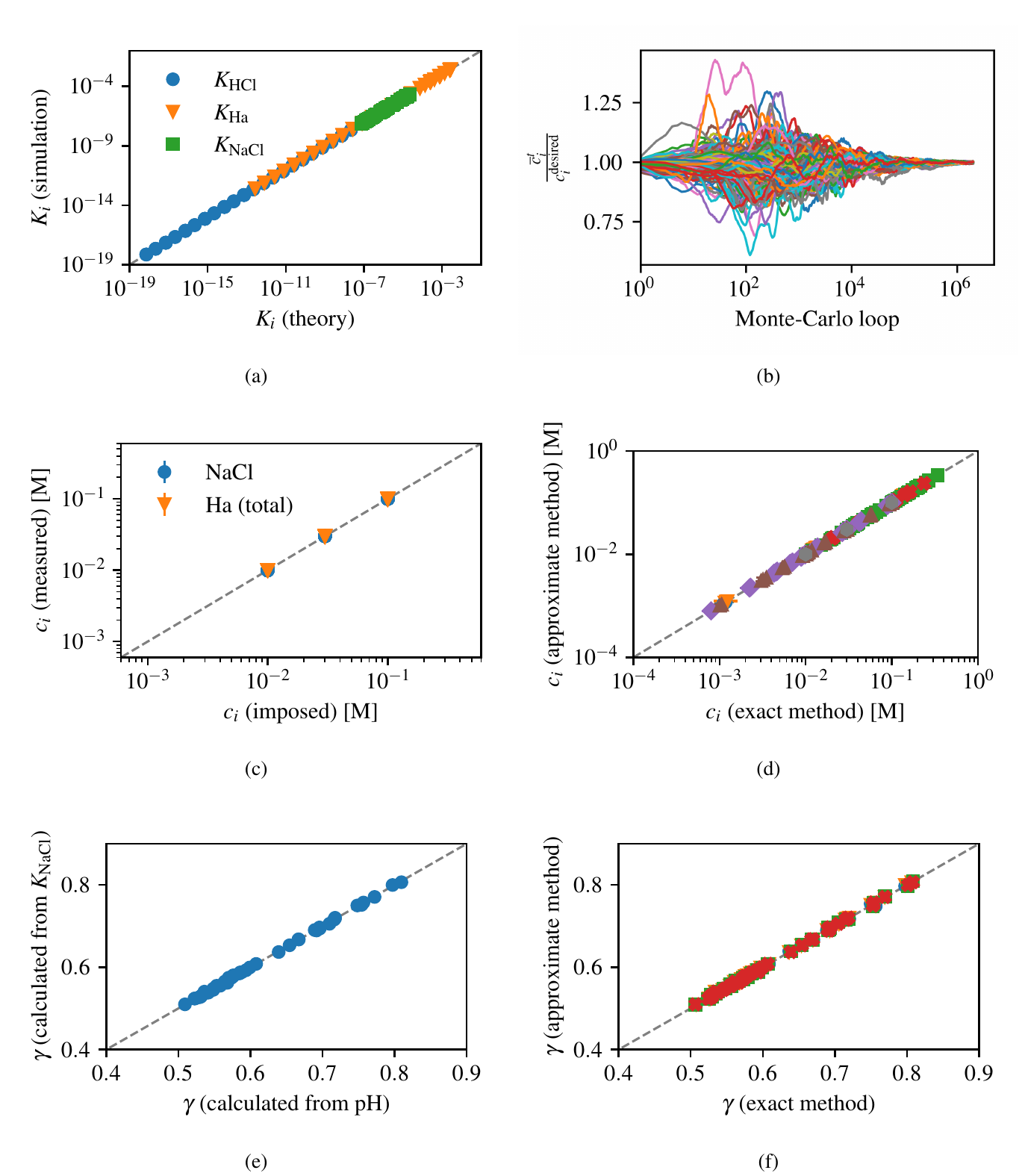}
\caption{\label{fig:esi_figure_1}(a): Plot of the various equilibrium constants obtained from the simulation of an ideal system vs. the values predicted by the ideal gas theory.
(b): Evolution of the mean concentrations $\overline{c_{\mathrm{NaCl}}}^t$ and $\overline{c_{\ce{Ha}}^{0}}^{t}$ for an interacting system for all simulated parameter combinations. To make the different simulations comparable, all concentrations are normalized by the respective desired concentration.
(c): Mean concentrations of the various species measured from the simulation for an interacting system (at the very end) vs. the concentrations imposed in the dynamical $\mu$-tuning algorithm.
(d):  Plot of the concentrations $c_i$ of the various species as obtained from the simulations using the approximate method vs the concentrations obtained from the simulations using the exact method.
(e): Plot of the activity coefficient $\gamma$ of an ion pair as calculated from $K_{\ce{NaCl}}$ vs $\gamma$ as calculated from the pH. The shown results were obtained using the approximate method.
(f): Plot of all possible combinations of $\gamma$ as calculated from either $K_{\ce{NaCl}}$ or $\gamma$ as calculated from the pH for the approximate and exact methods.}
\end{figure*}

Supplementary to the plot of the concentrations as measured from the simulations vs. the concentrations as predicted by the ideal gas theory provided in the main text (\autoref{fig:main_figure_2}), we show here an analogous plot for the equilibrium constants (\autoref{fig:esi_figure_1} (a)) where we observe again a very good agreement.

\subsubsection{Interacting System}
For the interacting case of a monoprotic acid, we show in \autoref{fig:esi_figure_1} (b) the time evolution of the mean concentrations $\overline{c_{\mathrm{NaCl}}}^t$ and $\overline{c_{\ce{Ha}}^{0}}^{t}$ for all simulated parameter combinations. Again (as in \autoref{fig:main_figure_2}), we normalize the concentrations by the respective desired concentrations in order to show all curves in the same plot.
We also show the mean concentrations of the various species measured from the simulation vs. the concentrations imposed in the dynamical $\mu$-tuning algorithm in \autoref{fig:esi_figure_1} (c).
As the plots show, convergence is reached for all simulations.

As an additional test, we compared the simulation results for this interacting system with a semi-analytical calculation that uses data from Widom insertion\cite{widom63a} simulations for a monovalent salt solution (compare \autoref{fig:main_figure_2} (c) in the main text).
The semi-analytical calculations were performed in the following way:
Neglecting the excluded volume effect of the neutral acid particle $\ce{Ha}$, which is a good approximation in the considered concentration regime, the composition of the reservoir for fixed input values of $\mathrm{pH}^\mathrm{res}$, $c_{\ce{NaCl}}^\mathrm{res}$ and $c_{\ce{Ha}}^{\mathrm{res},0}$ can be determined by solving the following system of non-linear equations:
\begin{align}
\mathrm{pH}^\mathrm{res} &= -\log_{10}\left(\frac{c_{\ce{H+}}^\mathrm{res}}{\cref}\sqrt{\gamma^\mathrm{res}}\right)\\
\gamma^\mathrm{res} &= f\left(I^\mathrm{res}\left(c_{\ce{H+}}^\mathrm{res},c_{\ce{OH-}}^\mathrm{res},c_{\ce{NaCl}}^\mathrm{res}, c_{\ce{a-}}^{\mathrm{res}}\right)\right)\\
c_{\ce{OH-}}^\mathrm{res} &= \frac{K_{\ce{H+},\ce{OH-}}(\cref)^2}{c_{\ce{H+}}^\mathrm{res}\gamma^\mathrm{res}}\\
c_{\ce{a-}}^{\mathrm{res}} &= \frac{c_{\ce{Ha}}^{\mathrm{res},0}}{1+\frac{\gamma^\mathrm{res} c_{\ce{H+}}^{\mathrm{res}}}{\cref K_{\mathrm{a}}}}.
\end{align}
In our calculation, we use linearly interpolated data from Widom insertion calculations for the activity coefficient $\gamma^\mathrm{res} = f(I)$.
Then, the above system of equations can be solved numerically, which we do in a self-consistent manner.

To further validate our method, we also performed some additional checks.
In \autoref{fig:esi_figure_1} (d), we plot the concentrations $c_i$ of all considered species as obtained from the simulations using the approximate scheme as described in \autoref{sec:det_re_c} vs the concentrations obtained using the exact method (which only works for the special case of a monoprotic acid, i.e. $n=1$)
As the plot shows, they are in agreement.
In \autoref{fig:esi_figure_1} (e), we show for the approximate method the activity coefficient of an ion pair as calculated from the equilibrium constant $K_{\ce{NaCl}}$ vs. the same activity coefficient as calculated from the pH. 
The fact that they agree shows that the results are internally consistent. 
Finally, in \autoref{fig:esi_figure_1} (f), we show a plot of all four possible combinations of $\gamma^{\text{res}}$ as calculated from the approximate and exact method using either $K_{\ce{NaCl}}$ or pH. 
The plot demonstrates that the two approaches yield the same results.

\subsection{Diprotic Acid}
In addition to the test for a monoprotic acid presented in the main text, we also ran tests for a diprotic acid. 
We retained the same simulation model and setup as above and ran tests for p$K_{\mathrm{a}}^1 = 4.0$,  p$K_{\mathrm{a}}^2 = 7.0$, $c_{\mathrm{NaCl}}^\mathrm{res}\in\left\lbrace 0.1\,\mathrm{M}, 0.03\,\mathrm{M}, 0.01\,\mathrm{M}\right\rbrace$, $c_{\ce{H2a}}^{\mathrm{res},0}\in\left\lbrace 0.1\,\mathrm{M}, 0.03\,\mathrm{M}, 0.01\,\mathrm{M}\right\rbrace$ and $\mathrm{pH}^\mathrm{res}\in\left[1.0,13.0\right]$.

\subsubsection{Ideal System}
\begin{figure*}[htb]
\centering
\includegraphics[width=0.8\textwidth]{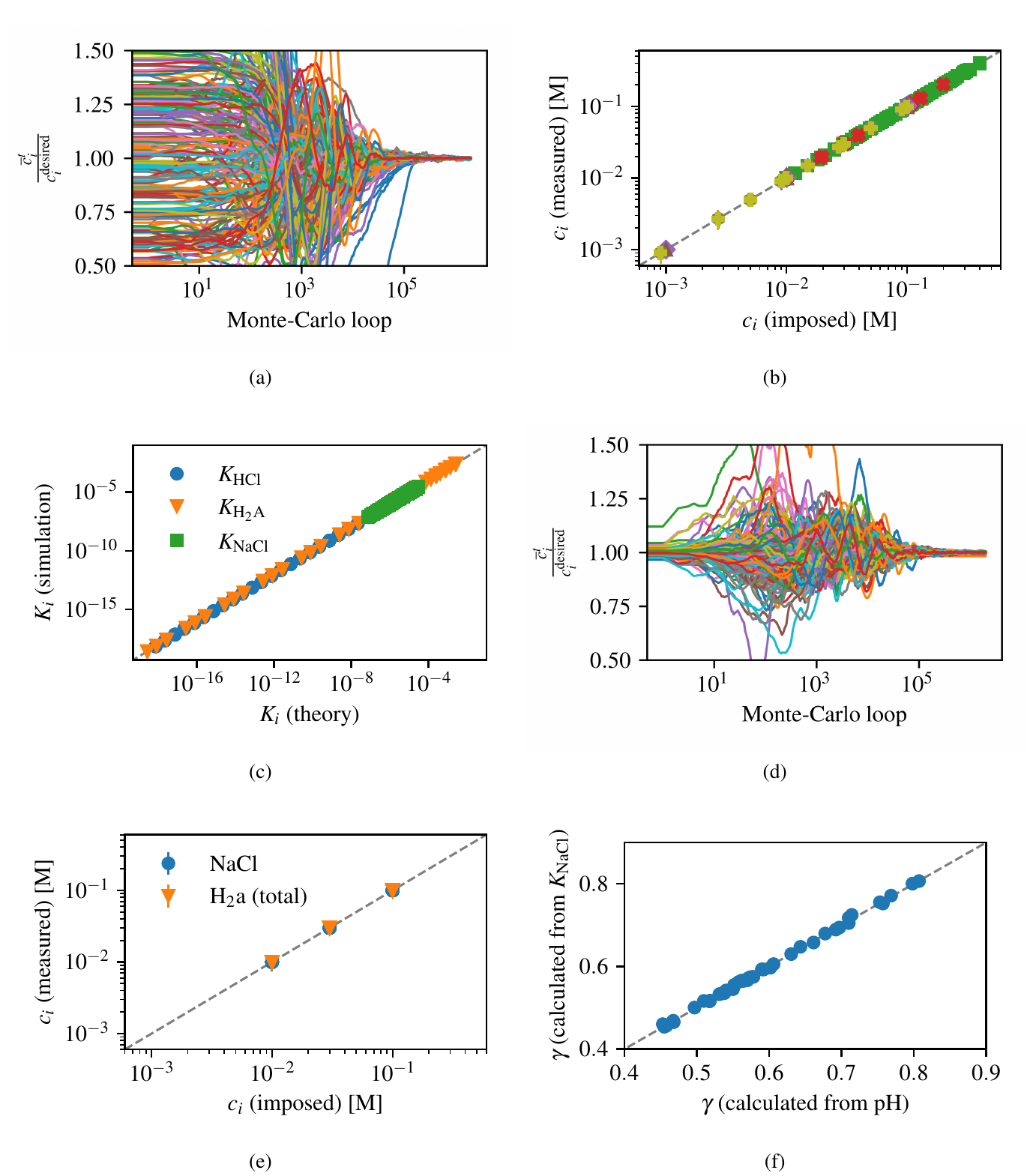}
	\caption{\label{fig:esi_figure_2}(a): Evolution of the mean concentrations $\overline{c_{\mathrm{NaCl}}}^t$ and $\overline{c_{\ce{Ha}}^{0}}^{t}$ for an ideal system containing a diprotic acid and for all simulated parameter combinations. To make the different simulations comparable, all concentrations are normalized by the respective desired concentration. 
(b): Mean concentrations of the various species measured from the simulation (at the very end) for an ideal system containing a diprotic acid vs. the concentration predicted from the ideal gas theory.
(c): Plot of the various equilibrium constants obtained from the simulation of an ideal system containing a diprotic acid vs. the values predicted by the ideal gas theory.
(d): Evolution of the mean concentrations $\overline{c_{\mathrm{NaCl}}}^t$ and $\overline{c_{\ce{H2a}}^{0}}^{t}$ for an interacting system containing a weak diprotic acid for all simulated parameter combinations. To make the different simulations comparable, all concentrations are normalized by the respective desired concentration.
(e): Mean concentrations of the various species measured from the simulation (at the very end) for an interacting system containing a diprotic acid vs. the concentrations imposed in the dynamical $\mu$-tuning algorithm.
(f): Plot of the activity coefficient $\gamma$ of a monovalent ion pair as calculated from $K_{\ce{NaCl}}$ vs $\gamma$ as calculated from the pH.
}
\end{figure*}

The first test case is again the ideal gas, which yields the following analytical solution:
\begin{align}
 c_{\ce{H+}}^{\mathrm{res}} &= \cref 10^{-\mathrm{pH}^\mathrm{res}}\\
 c_{\ce{OH-}}^{\mathrm{res}} &= \cref 10^{-\left(14-\mathrm{pH}^\mathrm{res}\right)}\\
 c_{\ce{H2a}}^{\mathrm{res}} &= \frac{c_{\ce{H2a}}^{\mathrm{res},0}}{1+\frac{\cref K_{\mathrm{a}}^1}{c_{\ce{H+}}^{\mathrm{res}}}+\frac{(\cref)^2 K_{\mathrm{a}}^1K_{\mathrm{a}}^2}{\left(c_{\ce{H+}}^{\mathrm{res}}\right)^2}}\\
 c_{\ce{Ha-}}^{\mathrm{res}} &= c_{\ce{H2a}}^{\mathrm{res}}\frac{\cref K_{\mathrm{a}}^1}{c_{\ce{H+}}^{\mathrm{res}}}\\
 c_{\ce{a^2-}}^{\mathrm{res}} &= c_{\ce{Ha-}}^{\mathrm{res}}\frac{\cref K_{\mathrm{a}}^2}{c_{\ce{H+}}^{\mathrm{res}}}\\
 c_{\ce{Na+}}^{\mathrm{res}} &= \max\left(c_{\ce{NaCl}}^{\mathrm{res}},c_{\ce{NaCl}}^{\mathrm{res}}+ c_{\ce{OH-}}^{\mathrm{res}}-c_{\ce{H+}}^{\mathrm{res}}+c_{\ce{Ha^-}}^{\mathrm{res}}+2c_{\ce{a^2-}}^{\mathrm{res}}\right)\\
 c_{\ce{Cl-}}^{\mathrm{res}} &= \max\left(c_{\ce{NaCl}}^{\mathrm{res}},c_{\ce{NaCl}}^{\mathrm{res}} - c_{\ce{OH-}}^{\mathrm{res}}+c_{\ce{H+}}^{\mathrm{res}}-c_{\ce{Ha^-}}^{\mathrm{res}}-2c_{\ce{a^2-}}^{\mathrm{res}}\right).
\end{align}
As a first plot, we show in \autoref{fig:esi_figure_2} (a) the evolution of the mean concentrations $\overline{c_{\mathrm{NaCl}}}^t$ and $\overline{c_{\ce{Ha}}^{0}}^{t}$ for an ideal system containing a diprotic acid and for all simulated parameter combinations. 
As before, to make the different simulations comparable, all concentrations are normalized by the respective desired concentration, showing the desired convergence.
In \autoref{fig:esi_figure_2} (b), we show again a plot of mean concentrations of the various species measured from the simulation (at the very end) vs. the concentration predicted from the ideal gas theory, which is also in excellent agreement.
We also observe a good agreement for the various equilibrium constants (\autoref{fig:esi_figure_2} (c)).

\subsubsection{Interacting System}
For the interacting version of this system, we show in \autoref{fig:esi_figure_2} (d) the evolution of the mean concentrations $\overline{c_{\mathrm{NaCl}}}^t$ and $\overline{c_{\ce{Ha}}^{0}}^{t}$ and observe again convergence.
In \autoref{fig:esi_figure_2} (e), we show again a plot of mean concentrations of the various species measured from the simulation (at the very end) vs. the imposed concentrations, which is also in agreement.
As a further consistency check, \autoref{fig:esi_figure_2} (f) shows the activity coefficient $\gamma$ of a monovalent ion pair as calculated from $K_{\ce{NaCl}}$ vs $\gamma$ as calculated from the pH.
Finally, in \autoref{fig:esi_figure_3} (a) we show the absolute value of the mean charge per acid particle as a function of the pH for the case $c_{\ce{NaCl}}^\text{res}=0.1$\,M and $c_{\ce{H2a}}^\text{res,0}=0.1$\,M.
As the plot shows, the ionization of the acid is enhanced as compared to the ideal result (Henderson-Hasselbalch equation) due to the interactions. 
This non-ideal effect is especially strong in the regime where the acid particles become ionized for the second time, i.e. when they become divalent ions.

\section{Additional Material for ``An Example System: Simulation of a Weak Polybase Solution Coupled to a Reservoir Containing a Weak Diprotic Acid''}

As mentioned in the main text, we performed also tests for an ideal realization of the polybase system described previously. 
In order to compare the results of these simulations to a theoretical prediction, we need to extend the ideal Donnan theory for a system that can exchange both monovalent and divalent ions with a reservoir.
The Donnan equilibrium is characterized by an equality of the chemical potentials of monovalent ion pairs $(+,-)$ and divalent ion triplets $(+,+,2-)$ between the two phases, i.e. 
\begin{align}
\mu_+^{\text{sys}} + \mu_-^{\text{sys}} &= \mu_+^{\text{res}} + \mu_-^{\text{res}}\\
2\mu_+^{\text{sys}} + \mu_{2-}^{\text{sys}} &= 2\mu_+^{\text{res}} + \mu_{2-}^{\text{res}}.
\end{align}
For an ideal gas, these conditions lead to the following equations for the concentrations of the different ions:
\begin{align}
 c_+^{\text{sys}}c_-^{\text{sys}} &= c_+^{\text{res}}c_-^{\text{res}}\\
 \left(c_+^{\text{sys}}\right)^2c_{2-}^{\text{sys}} &= \left(c_+^{\text{res}}\right)^2c_{2-}^{\text{res}}.
\end{align}
From these conditions, it is straightforward to see that the partition coefficients $\xi_i\equiv c_i^{\text{sys}}/c_i^{\text{res}}$ have to obey the following relation:
\begin{align}
 \xi_+ = \frac{1}{\xi_-} = \frac{1}{\sqrt{\xi_{2-}}}.
	\label{eq:partition_coefficients}
\end{align}
Combining this relation with the electroneutrality condition for the system,
\begin{align}
 c_{\ce{B+}}^{\text{sys}} + c_{\ce{H+}}^{\text{sys}} + c_{\ce{Na+}}^{\text{sys}} = c_{\ce{OH-}}^{\text{sys}} + c_{\ce{Cl-}}^{\text{sys}} + c_{\ce{Ha-}}^{\text{sys}} + 2c_{\ce{a^2-}}^{\text{sys}},
\end{align}
results in the non-linear equation 
\begin{align}
 c_{\ce{BH+}}^{\text{sys}} + \xi_+\left(c_{\ce{H+}}^{\text{res}} + c_{\ce{Na+}}^{\text{res}}\right) = \frac{1}{\xi_+}\left(c_{\ce{OH-}}^{\text{res}} + c_{\ce{Cl-}}^{\text{res}} + c_{\ce{Ha-}}^{\text{res}}\right) + \frac{2}{\xi_+^2}c_{\ce{a^2-}}^{\text{res}}
	\label{eq:nonlinear}
\end{align}
for the partition coefficient $\xi_+$, which can be brought into the form of a third-degree polynomial equation:
\begin{align}
	&\xi_+^3\left(c_{\ce{H+}}^{\text{res}} + c_{\ce{Na+}}^{\text{res}}\right) +\xi_+^2 c_{\ce{BH+}}^{\text{sys}}\\
&-\xi_+\left(c_{\ce{OH-}}^{\text{res}} + c_{\ce{Cl-}}^{\text{res}} + c_{\ce{Ha-}}^{\text{res}}\right) - 2c_{\ce{a^2-}}^{\text{res}} = 0.
\end{align}
In the case of a weak polybase, $c_{\ce{BH+}}^{\text{sys}}$ is not an independent parameter, but determined by the Henderson-Hasselbalch equation according to
\begin{align}
c_{\ce{BH+}}^{\text{sys}} = \frac{c_{\ce{B}}^{\text{sys},0}}{1+ 10^{\text{pH}^{\text{sys}}-\text{p}K_{\text{A}}^{\text{base}}}},
	\label{eq:HHbase}
\end{align}
where it is important to realize that $\text{pH}^{\text{sys}}\neq\text{pH}^{\text{res}}$ because of the Donnan potential and thus
\begin{align}
\begin{split}
\text{pH}^{\text{sys}} &= -\log_{10}\left(\frac{c_{\ce{H+}}^{\text{sys}}}{\cref}\right) = -\log_{10}\left(\frac{c_{\ce{H+}}^{\text{res}}}{\cref}\right) -\log_{10}\left(\frac{c_{\ce{H+}}^{\text{sys}}}{c_{\ce{H+}}^{\text{res}}}\right)\\
&= \text{pH}^{\text{res}} - \log_{10}\left(\xi_+\right).
\label{eq:pH}
\end{split}
\end{align}
This means that the Donnan partitioning and the ionization equilibrium of the weak polybase are mutually coupled. 
One might think that this coupling also has to be explicitly taken into account for the weak diprotic acid, however an inspection of \autoref{eq:partition_coefficients} and \autoref{eq:pH} reveals that the partitioning in combination with the shift in pH "automatically" leads to the correct behaviour.
Thus, one simply has to (numerically) solve \autoref{eq:nonlinear}, \autoref{eq:HHbase} and \autoref{eq:pH} in order to arrive at the correct value of $\xi_+$. 

\begin{figure}[htb]
\centering
\includegraphics[width=0.8\columnwidth]{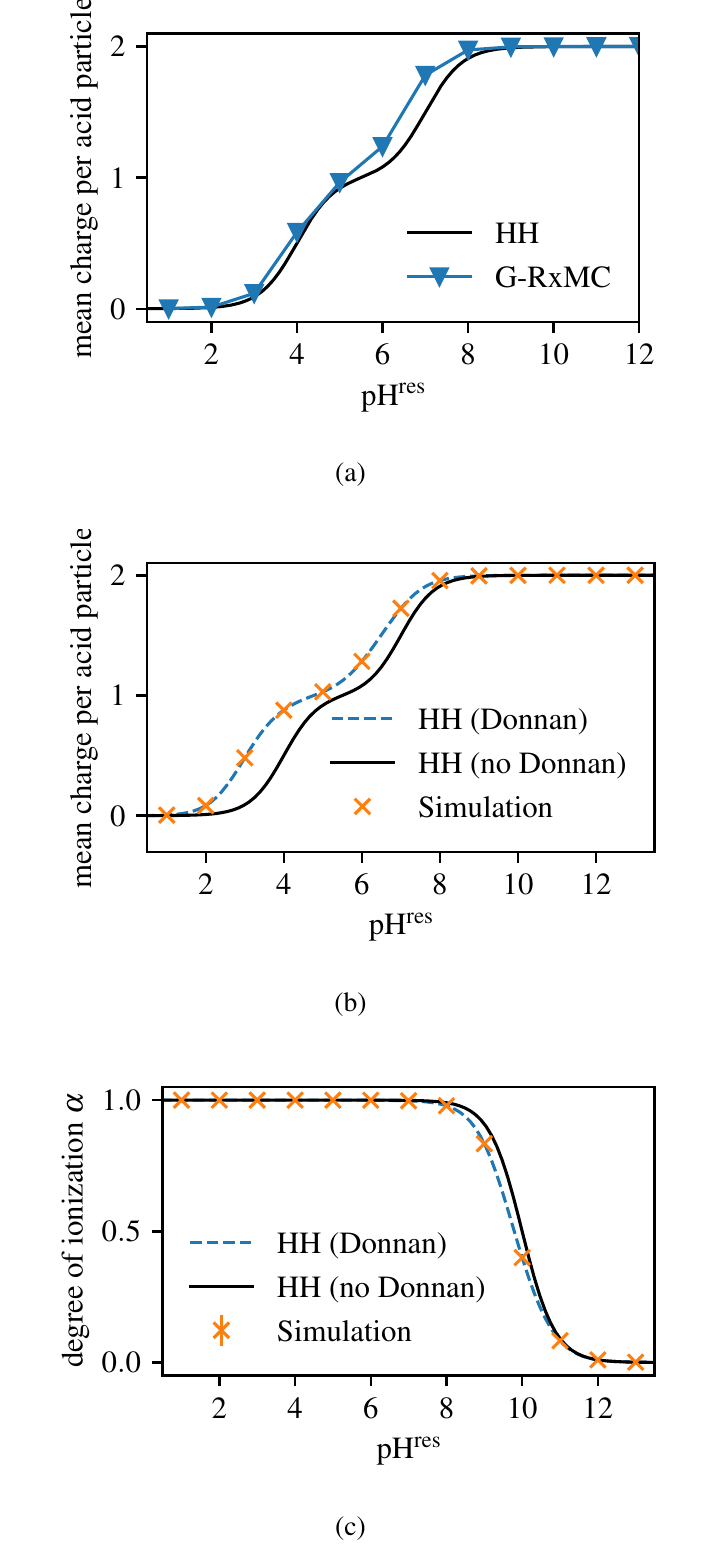}
  \caption{\label{fig:esi_figure_3}(a): Plot of the absolute value of the mean charge per diprotic acid particle for the interacting system vs the pH (data points). The shown plot corresponds to $c_{\ce{NaCl}}^\text{res}=0.1$\,M and $c_{\ce{H2a}}^\text{res,0}=0.1$\,M. As a comparison, the prediction by the ideal theory (Henderson-Hasselbalch equation) is also shown.
  (b): Plot of the absolute value of the mean charge per acid particle as a function of pH$^\mathrm{res}$. The predictions according to the ideal theory (Henderson-Hasselbalch equation), both with and without the Donnan effect are shown as well.
  (c): Plot of the degree of ionization of the weak polybase as a function of pH$^\mathrm{res}$.
  The predictions according to the ideal theory (Henderson-Hasselbalch equation), both with and without the Donnan effect are shown as well.
}
\end{figure}

In \autoref{fig:esi_figure_3} we show the results for this analytical approach in comparison to simulation results for an ideal gas.
As parameters, we chose p$K_\text{A}^1=4.0$, p$K_\text{A}^2=7.0$ and p$K_\text{A}^\text{base}=10.0$ as in the main text and set $c_{\mathrm{NaCl}}^\mathrm{res}= 0.1\,\mathrm{M}$, $c_{\ce{H2a}}^{\mathrm{res},0}=0.1\,\mathrm{M}$, $\mathrm{pH}^\mathrm{res}\in\left[1.0,13.0\right]$ with a monomer concentration of $c_{\ce{mon}}^{\mathrm{res},0}= 1.0\,\mathrm{M}$.
\autoref{fig:esi_figure_3} (b) shows the absolute value of the mean charge per acid particle as a function of $\mathrm{pH}^\mathrm{res}$.
The charge of the diprotic acid is strongly enhanced as compared to the naive prediction using the Henderson-Hasselbalch equation without the Donnan partitioning. 
This effect is a consequence of the Donnan effect, which effectively increases the pH inside the system as compared to $\mathrm{pH}^\mathrm{res}$.
As the plot shows, the simulation and the theoretical prediction are in quantiative agreement.
Similarly, we observe a perfect agreement for the degree of ionization of the weak polybase monomers, shown in \autoref{fig:esi_figure_3} (c).
In contrast to the acid, for the polybase the ionization is suppressed due to the Donann effect.

\section{Welford Method for the Covariance}
Miles et al.\cite{miles22a} showed how the Welford algorithm\cite{welford62a} to incrementally update the variance can be adapted to the case where one wants to incrementally update the variance of only the more recent half of a sample.
Because in our method we need not only to calculate the variance but also covariances, we show how this result generalizes to the covariance.
Similar to the case of the variance, let us write the covariance in the form
\begin{align}
\overline{\mathrm{Cov}}_t\lbrack x,y\rbrack = C_t / L_t
\end{align}
with
\begin{align}
C_t = \sum_{t'=\lceil ct\rceil}^{t} (x_{t'}-\overline{x}_t)(y_{t'}-\overline{y}_t).
\end{align}
As in the case of the variance, there are two cases to consider:
\begin{align}
&\text{Case A:}\quad \lceil c(t+1)\rceil=\lceil ct\rceil\\
&\text{Case B:}\quad \lceil c(t+1)\rceil=\lceil ct\rceil +1
\end{align}
A straightforward calculation (see below) shows that 
\begin{widetext}
\begin{align}
\label{eq:generalized_welford}
C_{t+1} = \begin{cases}
C_t + (x_{t+1}-\overline{x}_{t})(y_{t+1}-\overline{y}_{t+1})\, &\text{for Case A}\\
C_t + y_{t+1}\left(x_{t+1}-\overline{x}_{t+1}\right) + \overline{y}_{t}\left(x_{\lceil ct\rceil}-\overline{x}_{t+1}\right) + y_{\lceil ct\rceil}\left(\overline{x}_{t+1}-x_{\lceil ct\rceil}\right)\, &\text{for Case B}
\end{cases}
\end{align}
It is easy to see that this result reduces to the expression found for the variance by Miles et al.\cite{miles22a} in the special case $x=y$. In the following, we also provide the full calculation of this result. 
For Case A, we have
\begin{align*}
C_{t+1} - C_{t} = & \sum_{t'=\lceil c(t+1)\rceil}^{t+1} (x_{t'}-\overline{x}_{t+1})(y_{t'}-\overline{y}_{t+1})-\sum_{t'=\lceil ct\rceil}^{t} (x_{t'}-\overline{x}_t)(y_{t'}-\overline{y}_t)\\
=& \sum_{t'=\lceil ct\rceil}^{t+1} (x_{t'}-\overline{x}_{t+1})(y_{t'}-\overline{y}_{t+1})-\sum_{t'=\lceil ct\rceil}^{t} (x_{t'}-\overline{x}_t)(y_{t'}-\overline{y}_t)\\
=& \text{ } (x_{t+1}-\overline{x}_{t+1})(y_{t+1}-\overline{y}_{t+1}) + \sum_{t'=\lceil ct\rceil}^{t}\left\lbrace(x_{t'}-\overline{x}_{t+1})(y_{t'}-\overline{y}_{t+1})-(x_{t'}-\overline{x}_t)(y_{t'}-\overline{y}_t)\right\rbrace\\
=& \text{ } (x_{t+1}-\overline{x}_{t+1})(y_{t+1}-\overline{y}_{t+1}) \text{ }+ \\
& +\sum_{t'=\lceil ct\rceil}^{t}\left\lbrace -\overline{x}_{t+1}y_{t'} -x_{t'}\overline{y}_{t+1} + \overline{x}_{t+1}\overline{y}_{t+1} + x_{t'}\overline{y}_t + \overline{x}_ty_{t'} - \overline{y}_t\overline{x}_t \right\rbrace\\
=& \text{ } (x_{t+1}-\overline{x}_{t+1})(y_{t+1}-\overline{y}_{t+1}) + \left(\overline{x}_{t+1}\overline{y}_{t+1}- \overline{y}_t\overline{x}_t \right)\sum_{t'=\lceil ct\rceil}^{t}1\text{ }+\\
&+\text{ }\left(\overline{x}_t-\overline{x}_{t+1}\right)\sum_{t'=\lceil ct\rceil}^{t} y_{t'} +\left(\overline{y}_t-\overline{y}_{t+1}\right)\sum_{t'=\lceil ct\rceil}^{t} x_{t'}\\
=& \text{ } (x_{t+1}-\overline{x}_{t+1})(y_{t+1}-\overline{y}_{t+1}) + \left(\overline{x}_{t+1}\overline{y}_{t+1}- \overline{y}_t\overline{x}_t \right)L_t +\left(\overline{x}_t-\overline{x}_{t+1}\right)L_t\overline{y}_t+\left(\overline{y}_t-\overline{y}_{t+1}\right)L_t\overline{x}_t\\
=& \text{ } (x_{t+1}-\overline{x}_{t+1})(y_{t+1}-\overline{y}_{t+1}) + L_t\left(\overline{x}_{t+1}\overline{y}_{t+1}-\overline{x}_{t+1}\overline{y}_t + \overline{y}_t\overline{x}_t - \overline{y}_{t+1}\overline{x}_t\right)\\
=& \text{ } (x_{t+1}-\overline{x}_{t+1})(y_{t+1}-\overline{y}_{t+1}) + L_t\left(\overline{x}_{t+1} - \overline{x}_{t}\right)\left(\overline{y}_{t+1} - \overline{y}_{t}\right)\\
=& \text{ } (x_{t+1}-\overline{x}_{t+1})(y_{t+1}-\overline{y}_{t+1}) + L_t\left(\overline{x}_{t+1} - \overline{x}_{t}\right)\left(\overline{y}_{t+1} - \frac{L_t+1}{L_t}\left(\overline{y}_{t+1}-\frac{y_{t+1}}{L_t+1}\right)\right)\\
=& \text{ } (x_{t+1}-\overline{x}_{t+1})(y_{t+1}-\overline{y}_{t+1}) + \left(\overline{x}_{t+1} - \overline{x}_{t}\right)\left(y_{t+1}-\overline{y}_{t+1}\right)\\
=& \text{ } (x_{t+1}-\overline{x}_{t})(y_{t+1}-\overline{y}_{t+1}).
\end{align*}
For Case B, we have
\begin{align*}
C_{t+1} - C_{t} = & \sum_{t'=\lceil c(t+1)\rceil}^{t+1} (x_{t'}-\overline{x}_{t+1})(y_{t'}-\overline{y}_{t+1})-\sum_{t'=\lceil ct\rceil}^{t} (x_{t'}-\overline{x}_t)(y_{t'}-\overline{y}_t)\\
= & \sum_{t'=\lceil ct\rceil+1}^{t+1} (x_{t'}-\overline{x}_{t+1})(y_{t'}-\overline{y}_{t+1})-\sum_{t'=\lceil ct\rceil}^{t} (x_{t'}-\overline{x}_t)(y_{t'}-\overline{y}_t)\\
= & \text{ }(x_{t+1}-\overline{x}_t)(y_{t+1}-\overline{y}_t) - (x_{\lceil ct\rceil}-\overline{x}_t)(y_{\lceil ct\rceil}-\overline{y}_t) +\text{ }\\
&+ \sum_{t'=\lceil ct\rceil+1}^{t+1} \left\lbrace (x_{t'}-\overline{x}_{t+1})(y_{t'}-\overline{y}_{t+1})-(x_{t'}-\overline{x}_t)(y_{t'}-\overline{y}_t)\right\rbrace\\
= & \text{ }(x_{t+1}-\overline{x}_t)(y_{t+1}-\overline{y}_t) - (x_{\lceil ct\rceil}-\overline{x}_t)(y_{\lceil ct\rceil}-\overline{y}_t) +\text{ }\\
&+ \sum_{t'=\lceil ct\rceil+1}^{t+1} \left\lbrace -\overline{x}_{t+1}y_{t'} -x_{t'}\overline{y}_{t+1} + \overline{x}_{t+1}\overline{y}_{t+1} + x_{t'}\overline{y}_t + \overline{x}_ty_{t'} - \overline{y}_t\overline{x}_t \right\rbrace\\
= & \text{ }(x_{t+1}-\overline{x}_t)(y_{t+1}-\overline{y}_t) - (x_{\lceil ct\rceil}-\overline{x}_t)(y_{\lceil ct\rceil}-\overline{y}_t) +\text{ }\\
&+ \left(\overline{x}_{t+1}\overline{y}_{t+1}- \overline{y}_t\overline{x}_t \right)\sum_{t'=\lceil ct\rceil+1}^{t+1}1\text{ } +\text{ }\left(\overline{x}_t-\overline{x}_{t+1}\right)\sum_{t'=\lceil ct\rceil+1}^{t+1} y_{t'} +\left(\overline{y}_t-\overline{y}_{t+1}\right)\sum_{t'=\lceil ct\rceil+1}^{t+1} x_{t'}\\
= & \text{ }(x_{t+1}-\overline{x}_t)(y_{t+1}-\overline{y}_t) - (x_{\lceil ct\rceil}-\overline{x}_t)(y_{\lceil ct\rceil}-\overline{y}_t) +\text{ }\\
&+ \left(\overline{x}_{t+1}\overline{y}_{t+1}- \overline{y}_t\overline{x}_t \right)L_{t+1}\text{ } +\text{ }\left(\overline{x}_t\overline{y}_{t+1}-\overline{x}_{t+1}\overline{y}_{t+1}\right)L_{t+1}  +\left(\overline{y}_t\overline{x}_{t+1}-\overline{y}_{t+1}\overline{x}_{t+1}\right)L_{t+1}\\
= & \text{ }(x_{t+1}-\overline{x}_t)(y_{t+1}-\overline{y}_t) - (x_{\lceil ct\rceil}-\overline{x}_t)(y_{\lceil ct\rceil}-\overline{y}_t) +\text{ }\\
& +\text{ }\left(\overline{x}_{t}\overline{y}_{t+1}+\overline{x}_{t+1}\overline{y}_{t}-\overline{x}_{t+1}\overline{y}_{t+1}- \overline{y}_t\overline{x}_t \right)L_{t+1}\\
= & \text{ }(x_{t+1}-\overline{x}_t)(y_{t+1}-\overline{y}_t) - (x_{\lceil ct\rceil}-\overline{x}_t)(y_{\lceil ct\rceil}-\overline{y}_t) + \left(\overline{y}_{t+1}-\overline{y}_{t}\right)\left(\overline{x}_{t}-\overline{x}_{t+1}\right)L_{t+1}\\
= & \text{ }(x_{t+1}-\overline{x}_t)(y_{t+1}-\overline{y}_t) - (x_{\lceil ct\rceil}-\overline{x}_t)(y_{\lceil ct\rceil}-\overline{y}_t) +\text{ }\\
& +\text{ }\left(\overline{y}_{t+1}-\overline{y}_{t+1} + \frac{y_{t+1}-y_{\lceil ct\rceil}}{L_t}\right)\left(\overline{x}_{t}-\overline{x}_{t+1}\right)L_{t}\\
= & \text{ }(x_{t+1}-\overline{x}_t)(y_{t+1}-\overline{y}_t) - (x_{\lceil ct\rceil}-\overline{x}_t)(y_{\lceil ct\rceil}-\overline{y}_t) +\text{ }\left(y_{t+1}-y_{\lceil ct\rceil}\right)\left(\overline{x}_{t}-\overline{x}_{t+1}\right)\\
= &\text{ } y_{t+1}\left(x_{t+1}-\overline{x}_{t+1}\right) + \overline{y}_{t}\left(x_{\lceil ct\rceil}-\overline{x}_{t+1}\right) + y_{\lceil ct\rceil}\left(\overline{x}_{t+1}-x_{\lceil ct\rceil}\right).
\end{align*}
\end{widetext}

Which proves \autoref{eq:generalized_welford}.

\bibliography{bibtex/icp, additional_refs.bib}

\end{document}